\documentclass[10pt, aps, rmp, superscriptaddress, nofootinbib,
               amsmath, 
               twocolumn, preprintnumbers,
               raggedbottom,
               floatfix,
               longbibliography, noeprint
              ]{revtex4-1}
\pdfoutput=1
\usepackage[T1]{fontenc}
\usepackage[utf8]{inputenc}
\usepackage[secnoindent]{my_paper}      
\usepackage{my_acronyms}        
\usepackage[english]{babel}     
\usepackage{adjustbox}          

\newcommand{\grad}[1]{\vect{\nabla} #1}
\newcommand{\laplacian}[1]{\vect{\nabla}^2 #1}

\begin{document}

\title{Nuclear Energy Density Functionals: What do we really know?}

\author{Aurel Bulgac}%
\email{bulgac@uw.edu}%
\affiliation{Department of Physics,%
  University of Washington, Seattle, Washington 98195--1560, USA}

\author{Michael McNeil Forbes}%
\email{mforbes@alum.mit.edu}%
\affiliation{Department of Physics \& Astronomy,%
  Washington State University, Pullman, Washington 99164--2814, USA}%
\affiliation{Department of Physics,%
  University of Washington, Seattle, Washington 98195--1560, USA}

\author{Shi Jin}%
\email{kingstone1991@gmail.com}%
\affiliation{Department of Physics,%
  University of Washington, Seattle,
  Washington 98195--1560, USA}

\date{\today}
\preprint{NT@UW-15-05}


\begin{abstract}
  We present the simplest \gls{NEDF} to date, determined by only 4
  significant phenomenological parameters, yet capable of fitting
  measured nuclear masses with better accuracy than the
  Bethe-Weizsäcker mass formula, while also describing density
  structures (charge radii, neutron skins etc.) and time-dependent
  phenomena (induced fission, giant resonances, low energy nuclear
  collisions, etc.).  The 4 significant parameters are necessary to
  describe bulk nuclear properties (binding energies and charge
  radii); an additional 2 to 3 parameters have little influence on the
  bulk nuclear properties, but allow independent control of the
  density dependence of the symmetry energy, excitation energy
  of isovector excitations, and the Thomas-Reiche-Kuhn sum rule.  This
  Hohenberg-Kohn--style of \gls{DFT} successfully realizes
  Weizsäcker's ideas and provides a computationally tractable model
  for a variety of static nuclear properties and dynamics, from finite
  nuclei to neutron stars, where it will also provide a new insight
  into the physics of the r-process, nucleosynthesis, neutron star
  crust structure, and neutron star mergers.  This new \gls{NEDF} clearly separates the bulk
  geometric properties -- volume, surface, symmetry, and Coulomb
  energies which amount to $\sim\SI{8}{MeV}$ per nucleon or up to
  $\sim \SI{2000}{MeV}$ per nucleus for heavy nuclei -- from finer
  details related to shell effects, pairing, isospin breaking,
  etc\@. which contribute at most a few \si{MeV} for the entire
  nucleus.  Thus it provides a systematic framework for organizing
  various contributions to the \gls{NEDF}.  Measured and calculated
  physical observables -- i.e\@. symmetry and saturation properties,
  the neutron matter equation of state, and the frequency of giant
  dipole resonances -- lead directly to new terms, not considered in
  current \gls{NEDF} parameterizations.
\end{abstract}

\glsreset{NEDF}
\glsreset{GDR}

\maketitle

\tableofcontents

\section{Introduction}
Calculating nuclear masses, nuclear matter and charge distributions,
and dynamics in nuclear systems remains one of the most challenging
problems in quantum many-body theory.  Almost a century ago,
\textcite{Aston:1920} realized that a nucleus is not quite the sum of
its parts, leading \textcite{Eddington:1920} to correctly conjecture a
link between nuclear masses, the conversion of hydrogen into heavier
elements, and the energy radiated by the stars. When quantum mechanics
was first applied to many-body systems, \textcite{Weizsacker:1935}
proposed that an energy density approach could be effective for
calculating nuclear binding energies.  This was the first instance of
an energy density functional being applied in nuclear physics, with
the fundamentals of \gls{DFT} laid several decades
later~\cite{HK:1964, Kohn:1965fk,
  Dreizler:1990lr}. \textcite{Bethe:1936} further developed
Weizsäcker's ideas and introduced the nuclear mass formula (known as
the Bethe-Weizsäcker formula) for the
ground state energies of nuclei with $A=N+Z$ nucleons ($N$ neutrons
and $Z$ protons):
\begin{equation}\label{eq:Bethe}
  E(N, Z) = a_v A + a_s A^{2/3} + a_C\frac{Z^2}{A^{1/3}}
  + a_I\frac{(N-Z)^2}{A}.
\end{equation}
Unlike electrons in atoms, nuclei are saturating systems with a nearly
constant interior density, yielding the terms above: a volume energy,
a surface tension, a non-extensive Coulomb energy, and a symmetry
energy that favors similar numbers of protons and neutrons.  As shown
in the first row of Table~\ref{table:liquid_drop}, these four terms
alone fit the latest evaluated nuclear
masses~\cite{Audi:2012,Wang:2012}  with a
\gls{rms} error of $\chi_E \approx$~\SI{3}{MeV} per nucleus, where
$\chi_E^2 =\sum\abs{E_{N,Z} - E(N, Z)}^2/N_E$ and we fit the $N_E =
2375$ measured (not extrapolated) nuclear masses of nuclei with $A\geq
16$ from~\textcite{Audi:2012, Wang:2012}. This is a remarkable result: the nuclear
binding energy of heavy nuclei can reach \SI{2000}{MeV}, hence
the errors are at the sub-percent level.

A slightly better fit is obtained using a mass formula with surface
corrections terms to the symmetry and Coulomb energies, as well
as odd-even staggering (due to pairing):
\begin{subequations}
  \label{eq:masses}
  \begin{multline}
    E(N, Z) = a_v A + a_s A^{2/3} + a_C\frac{Z^2}{A^{1/3}}
    + a'_C\frac{Z^2}{A^{2/3}} \\
    + a_I\frac{(N - Z)^2}{A} + a'_I\frac{(N - Z)^2}{A^{4/3}}
    + \Delta.
  \end{multline}
  \begin{gather}
    \Delta = \begin{cases}
      -\delta A^{-1/2} & \text{even-even nuclei,} \\
      \hfil 0 & \text{odd nuclei,} \\
      \hfil\hphantom{-}\delta A^{-1/2} & \text{odd-odd nuclei.}
    \end{cases} \label{eq:odd-even}
  \end{gather}
\end{subequations}
This pairing contribution is significantly smaller than the others,
with an amplitude $\approx \SI{12}{MeV}/A^{1/2}$; it is also smaller
than contributions arising from shell-correction energies, changing
the rms error $\chi_E$ by about at most \SIrange{100}{150}{keV}.  The
results of this fit are shown in Table~\ref{table:liquid_drop} with
the overall residuals displayed in Fig.~\ref{fig:masses}.  The
magnitudes of the various terms are compared in
Fig.~\ref{fig:mass_contributions}. While the volume, surface, and
Coulomb contributions are clearly dominant to the nuclear binding
energy, the symmetry energy contributions are roughly at the level of
10\% at most.

\renewcommand{\th}[1]{\multicolumn{1}{c}{#1}}  
\newcolumntype{3}{D{.}{.}{3}}
\newcolumntype{2}{D{.}{.}{2}}
\newcolumntype{1}{D{.}{.}{1}}
\newcolumntype{i}{D{.}{.}{0}}
\begin{table}[t]
  \begin{ruledtabular}
    \begin{tabular}{22223322}
      \th{$a_v$} & \th{$a_s$} & \th{$a_I$} & \th{$a'_I$} & \th{$a_C$} &
      \th{$a'_C$} & \th{$\delta$} & \th{$\chi_E$} \\
      -15.46 & 16.71 & 22.84 &   0    & 0.698 &  0     &  0    & 3.30 \\
      -15.48 & 16.76 & 22.88 &   0    & 0.699 &  0     & 12.29 & 3.17 \\
      -15.31 & 17.74 & 24.94 & -22.50 & 0.765 & -0.667 &  0    & 2.64 \\
      -15.34 & 17.78 & 24.99 & -22.33 & 0.765 & -0.653 & 11.46 & 2.50
    \end{tabular}
  \end{ruledtabular}\label{table:liquid_drop}
  \caption{Parameters and the energy rms of the mass formulas
    Eqs.~\eqref{eq:Bethe} or~\eqref{eq:masses}, with or without the
    contribution Eq.~\eqref{eq:odd-even} for 2375 nuclei
    from \textcite{Audi:2012, Wang:2012}. (All quantities express in \si{MeV}.)
  }
\end{table}
\begin{figure}[t]
  \includegraphics[width=\columnwidth]{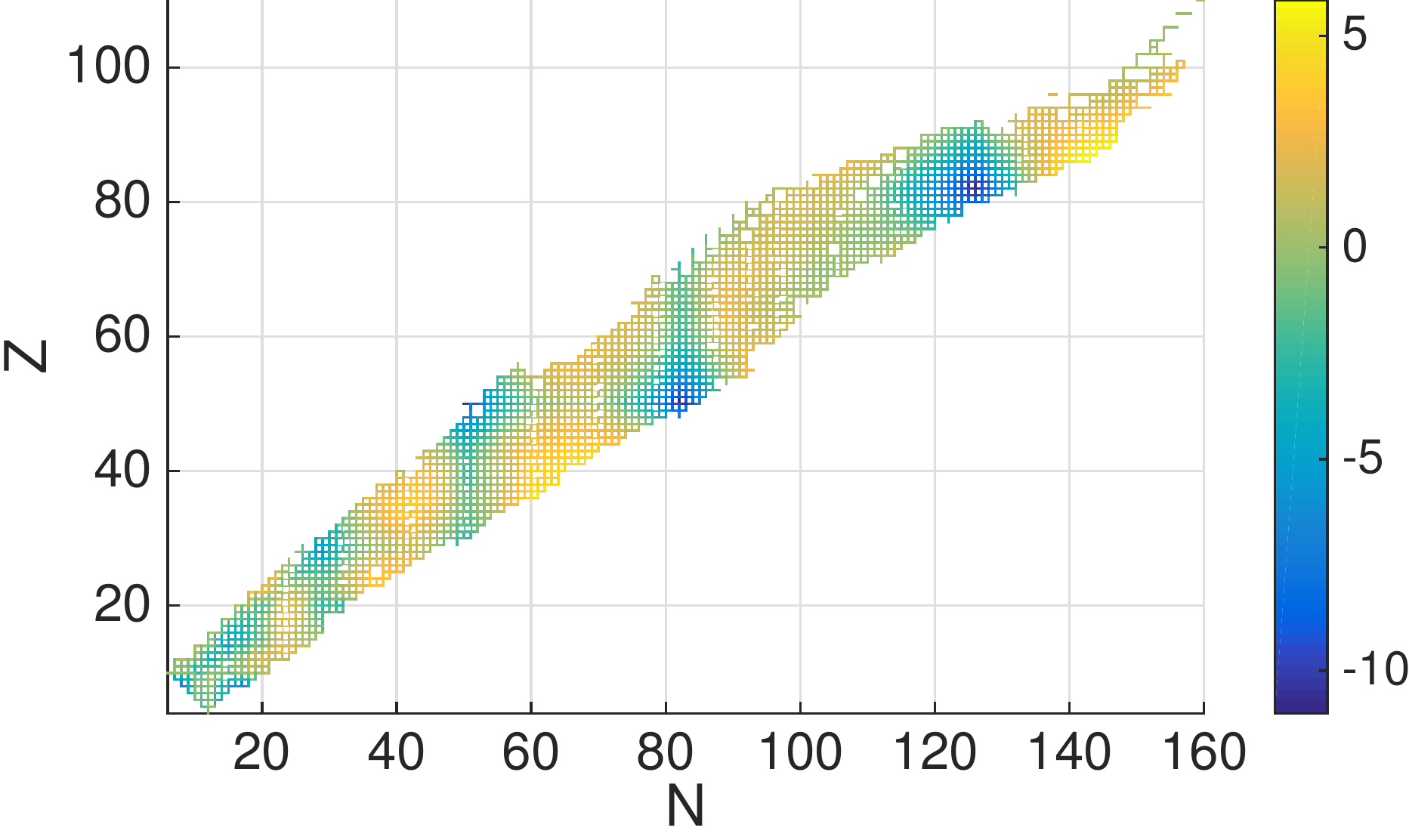}
  \caption{\label{fig:masses}%
    (Color online) The differences $E_{\text{exp}} - E_{\text{th}}$ in
    \si{MeV}s between the evaluated ground state
    energies~\cite{Audi:2012, Wang:2012} of 2375 nuclei with $A\ge 16$
    and fitted with the 6-parameter mass formula
    Eq.~\eqref{eq:masses} and $\Delta \equiv 0$. One can easily
    identify the emergence of closed shells for protons and neutrons.}
\end{figure}
\begin{figure}[t]
  \includegraphics[width=\columnwidth]{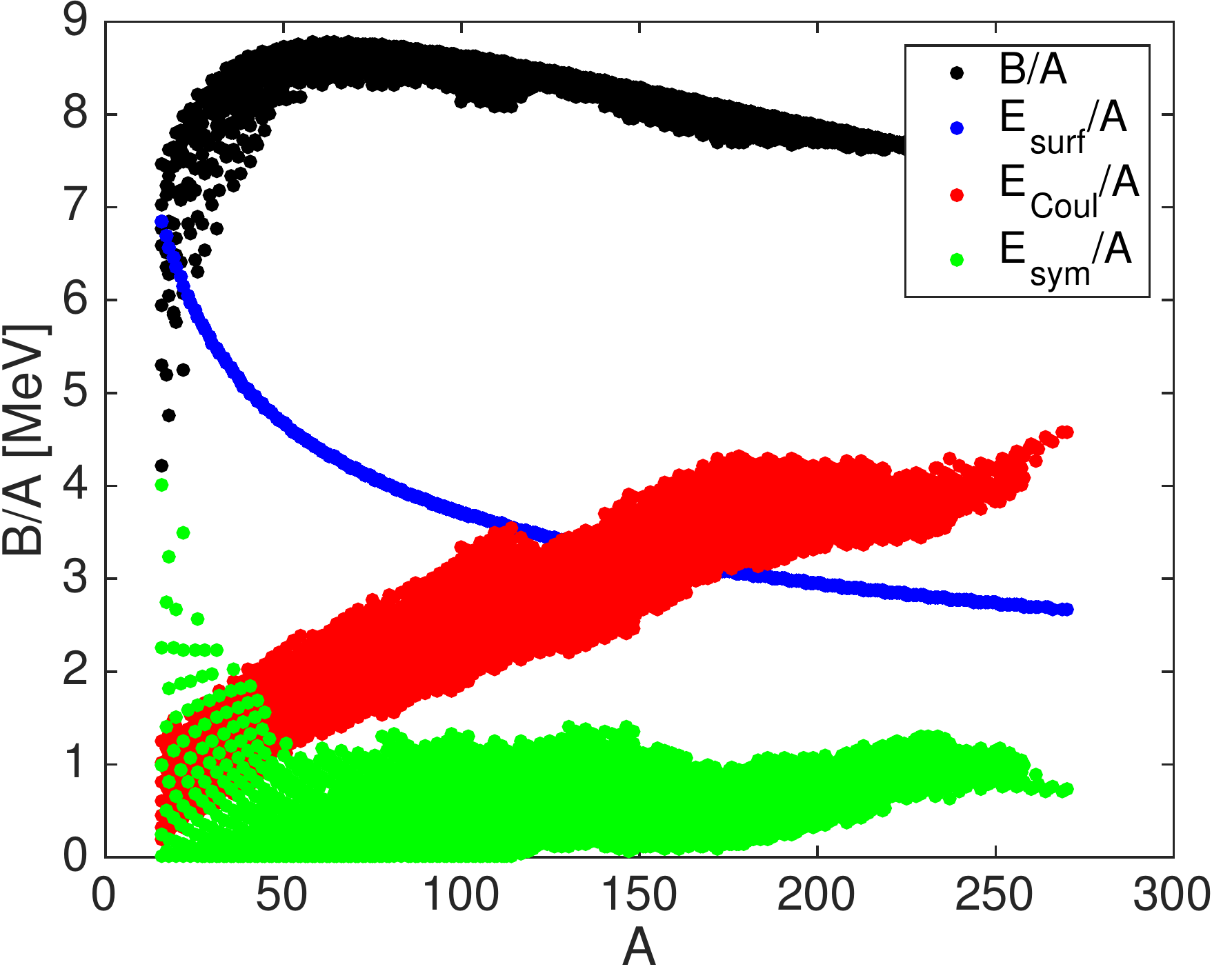}
  \caption{\label{fig:mass_contributions}%
    (Color online) The binding energy per nucleon $B/A = |E(N,Z)|/A$
    and the Coulomb, surface and symmetry energy per nucleon in
    Eq.~\eqref{eq:masses} for the measured 2375 nuclei with $A \ge
    16$~\cite{Audi:2012, Wang:2012}.}
\end{figure}

In parallel, a number of properties of many-fermion systems were
understood mathematically by tying together the roles of the
geometry and of the periodic trajectories in cavities.  As early as
1911, \textcite{Weyl:1911, Weyl:1912, Weyl:1912a, Weyl:1912b,
  Weyl:1913, Weyl:1915, Weyl:1950} and others related the
wave eigenstate density in boxes of various shapes and
boundary conditions to the geometrical shape of the
box~\cite{Kac:1966, Waechter:1972, Baltes:1976, Brack:1997}.  In a
manner very similar to the nuclear mass formula Eq.~\eqref{eq:Bethe},
such formulas can be applied to saturating systems, relating the
ground state energy to the volume ($V$), surface area ($A$), and mean
curvature radius $R$ of the many-particle system:
\begin{gather}\label{eq:Weyl}
  E = a_V V + a_S S + a_R R + \dots.
\end{gather}
The similarity to the nuclear mass formula Eq.~\eqref{eq:Bethe} becomes
apparent after relating the volume to the particle number $\rho = A/V\approx
\text{const}$.  (The Coulomb repulsion energy should to be evaluated separately in
a straightforward manner, due to the long-range character of the interaction.)
The ground state energy can thus be rewritten in terms of particle number $A$
(here for only one kind of particles)
\begin{gather}\label{eq:EA}
  E = b_V A + b_S A^{2/3} + b_R A^{1/3} + \dots.
\end{gather}
The coefficient $b_V$ is the energy per particle in infinite
matter and $a_S$ represents the surface tension.  These types of expansion
are essentially classical in character, with Planck's constant playing no
explicit role, and their accuracy for many-fermion system is limited by the
lack of quantum effects (referred often as shell effects).  It appears that
for nuclei the
mass formula Eq.~\eqref{eq:masses} is about as good as one can achieve without
introducing the quantum effects.

\textcite{Balian:1970, Balian:1971, Balian:1972} observed that quantum
states in a finite system can be quite accurately reproduced by
quantizing the periodic classical trajectories.  Combining the idea of
geometrical quantization with the Thomas-Fermi model, the Pauli
principle, and copious empirical evidence that strongly interacting
fermionic systems share many similarities with non-interacting
systems~\cite{Haxel:1949, Mayer:1949, Mayer:1950, Mayer:1950a,
  Landau:1956, Landau:1957, Migdal:1967}, one can quite accurately
construct the single-particle density of states and binding energies
as a function of the particle number, eventually correcting this by
the shape of the system.

The single-particle density of states $\rho(\epsilon)$ in a given potential
has a smooth and an oscillating components:
\begin{subequations}
  \begin{gather}
    \rho(\epsilon) = \rho_{\text{TF}}(\epsilon) + \rho_{\text{osc}}(\epsilon),
    \label{eq:rho_e}\\
    \rho_{\text{osc}}(\epsilon)= \sum_{\text{PO}}a_{\text{PO}}(\epsilon)
    \sin\left( \frac{S_{\text{PO}}(\epsilon)}{\hbar}+\phi_{\text{PO}}\right) +\dots,
  \end{gather}
\end{subequations}
where the sum is performed over classical periodic orbits (PO)
(diameter, triangles, squares, etc.), $a_{\text{PO}}(\epsilon)$ is
the stability amplitude, $S_{\text{PO}}(\epsilon)$ the action, and
$\phi_{\text{PO}}$ the Maslov index of each orbit at the energy
$\epsilon$~\cite{Balian:1970, Balian:1971, Balian:1972, Nishioka:1990,
  Brack:1997}.  The single-particle density of states in the
Thomas-Fermi approximation $\rho_{\text{TF}}$~\cite{Weyl:1911,
  Weyl:1912, Weyl:1912a, Weyl:1912b, Weyl:1913, Weyl:1915, Weyl:1950,
  Baltes:1976, Kac:1966, Waechter:1972,Brack:1997} has a clear
dependence on the size and shape of the system, and leads to
Eqs.~\eqref{eq:Weyl} and \eqref{eq:EA} for a square-well potential.
At the same time, the nature of the periodic orbits also depends on
the size and shape of the single-particle potential.

Knowing $\rho(\epsilon)$, one can calculate the particle number $A$ and
shell-corrections (SC) energy $E_{\text{SC}} = E - E_{\text{TF}}$ for a
many-fermion system by integrating up to the chemical potential $\mu$:
\begin{align}
  A&=\int_{-\infty}^\mu\rho(\epsilon) d\epsilon, &
  E_{\text{SC}}= \int_{-\infty}^\mu
  \epsilon \rho_{\text{osc}}(\epsilon)d\epsilon.
  \label{eq:shell}
\end{align}

The theory of periodic orbits and structure of these shell-corrections has been
studied extensively.  For example, in a 3-dimensional spherical cavity, quantum
effects can be reproduced by including only triangular and
square orbits~\cite{Balian:1970, Balian:1971, Balian:1972, Nishioka:1990,
  Brack:1997}. The emergence of magic numbers, and the role of the shapes of
many-fermion systems have been tested in theory and validated against
experimental results in fermion systems with up to \num{3000}
electrons~\cite{Heer:1993, Brack:1993, Pedersen:1991}.  In particular, in
atomic clusters the emergence of the super-shells has been predicted
theoretically~\cite{Nishioka:1990, Brack:1993, Bulgac:1993} and confirmed
experimentally~\cite{Heer:1993, Pedersen:1991}, though nuclei are too small to
exhibit the emergence of super-shells.

In nuclear physics a similar line of inquiry is encapsulated in the
method of shell-corrections, developed by Strutinsky~\cite{Strutinsky:1966,
Strutinsky:1967, Strutinsky:1968} and
  others~\cite{Strutinsky:1976, Brack:1972, Brack:1985, Myers:1966,
    Myers:1969, Myers:1974, Myers:1990, Myers:1991, Myers:1996,
    Ring:2004, Bohr:1998, Moller:1995, Moller:2012},  which shows that
$\rho(\epsilon)$ has a well defined dependence on the particle
number. The smooth part of the density of states is well described by
the Thomas-Fermi approximation (and alternatively by the smoothing
procedure introduced by Strutinsky), the leading terms of which are the
volume ($\sim A$), surface ($\sim A^{2/3}$), Coulomb ($\sim
Z^2/A^{1/3}$), and symmetry energy ($\sim (N-Z)^2/A$) contributions
encoded in the Bethe-Weizsäcker mass formula~\eqref{eq:Bethe}.
The oscillating part is dominated by the
nuclear shape and the shell effects from the periodic orbits, with an
amplitude
dependent on the particle number as $A^{1/6}$~\cite{Strutinsky:1976}.

This separation of $\rho(\epsilon)$ into the smooth and oscillating
parts~\eqref{eq:rho_e} is a general characteristic of the many fermion
systems.  Both the macroscopic-microscopic method~\cite{Strutinsky:1966,
Strutinsky:1967, Strutinsky:1968,
    Strutinsky:1976, Brack:1972, Brack:1985, Myers:1966, Myers:1969,
    Myers:1974, Myers:1990, Myers:1991, Myers:1996, Ring:2004,
    Bohr:1998, Moller:1995, Moller:2012}  and self-consistent
approaches~\cite{Bender:2003, Goriely:2009, Delaroche:2010, Goriely:2013,
    Goriely:2014, Kortelainen:2014} lead to the same conclusions
about the various contributions described above, and agree with
experimental data~\cite{Lunney:2003}.  This suggests the question: To
what order can one expand the density of states in powers of the
particle numbers and periodic orbits?

There is a reasonable consensus that, beyond the leading contributions
from the periodic orbits and shell-corrections, such an expansion
fails due to the effects of quantum chaos -- i.e\@. contributions from
classically chaotic trajectories through the many-body phase
space~\cite{Bohigas:1993}.  Stable periodic orbits provide the
strongest shell effects in quantum systems, evident for example in the
magic numbers (see e.g.~Fig.~\ref{fig:be_zn}). Unstable periodic
orbits also produce shell effects, but typically with smaller
weights. In contrast, chaotic orbits appear to produce irregular
oscillations in the single-particle density of states with a rather
small amplitude.  Various estimates suggest that chaotic fluctuations
appear at the level of \SI{0.5}{MeV} per nucleus~\cite{Bohigas:2002,
*Bohigas:2002E, Aberg:2002,
    Olofsson:2006, Olofsson:2008, Molinari:2004, Molinari:2006,
    Hirsch:2004, Hirsch:2005, Barea:2005}, noticeably smaller than
shell effects contributions due to periodic orbits and deformations, which are of
the order of several \si{MeV}s.

The effect of periodic orbits is not limited to finite systems: the
Casimir energy in quantum field theory~\cite{Casimir:1948,
    Klimchitskaya:2009},
critical phenomena~\cite{Fisher:1978, Hanke:1998}, and strongly interacting
infinite inhomogeneous systems, e.g\@. nuclear pasta phase in neutron
stars~\cite{Bulgac:2001y,*Bulgac:2002x,
    Magierski:2003, Magierski:2004x, Magierski:2002x, Bulgac:2005x,
    Yu:2000, Bulgac:2006x}, can also be explained and calculated to high precision by
evaluating the contributions from periodic orbits.  This method has become the standard for
evaluating the Casimir energy in a variety of fields~\cite{Bordag:2010, Rahi:2009,
    Graham:2014, Canaguier-Durand:2010, Schaden:2010}.

It is somewhat surprising that shell effects from periodic orbits appear at the
same level as deformation effects in the energy of nuclear systems.
Naïvely one might expect the
deformation energy to be controlled by the surface area of a saturating system,
and thus to contribute as a correction to the surface term in nuclear mass
formulas like Eqs.~\eqref{eq:Bethe} and~\eqref{eq:masses}. However, the
deformation energy in nuclei has a quantum nature, and is determined by a
delicate interplay between energy changes induced by the changes in surface
area and the shell effects.
A similar behavior has been observed in the case of atomic clusters with up to
3000 electrons~\cite{Bulgac:1993}, leading to the leveling of the peaks, which
one would otherwise expect in the absence of
deformation), leaving in place only the large negative
shell-corrections for the magic spherical systems, as seen in
Fig.~\ref{fig:be_zn} in case of nuclei.

The shape stability of a many-fermion system is controlled by the
single-particle level density at the Fermi level. In an open-shell
system this level density is high; the system can thus deform quite
easily and single-particle levels can rearrange until the level
density is low enough to render the system stable.  The stabilization
process of the nuclear deformation in the ground state is analogous to
the Jahn-Teller effect in polyatomic molecules~\cite{Jahn:1937}, where
the high degeneracy of the ground state is lifted by the deformation
of the system.  This mechanism leads to new ``magic numbers'' in
deformed systems as Strutinsky discussed in his seminal
papers~\cite{Strutinsky:1966, Strutinsky:1967, Strutinsky:1968}.  The
increase in surface area and the energy penalty incurred (deformation
energy) is canceled to a large extent by the shell-corrections (due to
periodic orbits in the deformed potential), unless the system is
``magic'' or ``semi-magic''. This cancellation between deformation
energy and shell effects suggests that open-shell systems should be
easier to deform than magic systems. This property is consistent with
the character of the residuals remaining after the fit of the nuclear
binding energies with Bethe-Weizsäcker formulas like
Eqs.~\eqref{eq:Bethe} and \eqref{eq:masses} and shown in
Fig.~\ref{fig:masses} (see also Fig.~\ref{fig:be_zn}). The largest
residuals appear as large (negative) spikes at the shell closures for
spherical nuclei with magic numbers of either protons or/and neutrons,
while the expected (positive) peaks in between magic numbers are
flattened.

Our goal is to generate a phenomenological \gls{NEDF} with the minimal
number of physically motivated parameters required to describe static
bulk nuclear properties.  This new \gls{NEDF} may also describe
nuclear dynamics in real time.

\glsreset{NEDF}
\section{Static Properties}

The lesson from our brief historical review is that, since nuclei are
saturating systems with a rather well defined saturation density, the
bulk of the nuclear binding energy should be fixed by the geometry of
the nuclei (volume, surface area, curvature radius) to sub-percent
accuracy.  As demonstrated in Table~\ref{table:liquid_drop}, the
accuracy of the mass formulas Eqs.~\eqref{eq:Bethe}
and~\eqref{eq:masses} -- which both lack shell effects, deformation,
spin-orbit effects, pairing, etc\@. -- suggests that such a \gls{NEDF}
should be capable of describing at a similar level of accuracy both
the nuclear binding energies, and the proton and neutron matter
density distribution.  Therefore, we might reasonably expect that a
\gls{NEDF} will also describe the nuclear charge radii, for which
there is a large amount of accumulated data~\cite{Angeli:2013}.
Quantum effects enter at the level of a few \si{MeV}s per nucleus,
reducing the rms energy error $\chi_E$ from around \SI{3}{MeV} to about
\SI{0.5}{MeV}~\cite{Moller:1995, Moller:2012}, and are most pronounced
for magic or semi-magic nuclei, see Fig.~\ref{fig:masses}.  The shell
corrections will be neglected in this round, but can be added through
the Strutinsky procedure~\cite{Strutinsky:1966, Strutinsky:1967,
  Strutinsky:1968, Strutinsky:1976, Brack:1972, Brack:1985,
  Myers:1966, Myers:1969, Myers:1974, Myers:1990, Myers:1991,
  Myers:1996, Ring:2004, Bohr:1998, Moller:1995, Moller:2012}.

A number of corrective terms might be considered to improve the
accuracy of the nuclear mass formulas Eqs.~\eqref{eq:Bethe}
and~\eqref{eq:masses}. For example, in the Coulomb term, one might
replace $Z^2$ with $Z(Z-1)$ to correctly count the number of proton
pairs, and one might add an additional term proportional to $Z$ to
account for the Coulomb exchange interaction and
screening~\cite{Bulgac:1996}. Motivated by Eq.~\eqref{eq:EA}, one
might also consider including terms proportional to $A^{1/3}$ and
$A^0$. The symmetry energy terms might also be ``corrected'' by
replacing $(N-Z)^2/4$ with $T(T+1)$ where $T=\abs{N-Z}/2$. Finally,
one might introduce an additional correction to account for the Wigner
energy $\propto \abs{N-Z}$, which appears as a cusp in the nuclear
binding energies as a function of $N-Z$ (basically only for nuclei
with $\abs{N-Z}\le 2$).

However, including these corrections lead to very small improvements
in the energy rms $\chi_E$ beyond the value \SI{2.64}{MeV} obtained
with the main terms of Eq.~\eqref{eq:masses}.  These corrections are
eclipsed by the shell effects as seen in Fig.~\ref{fig:masses}.  From
the nature of the residuals $E_{\text{exp}} - E_{\text{th}}$ in
Fig.~\ref{fig:masses} -- sharp negative spikes at the magic numbers, but roughly
constant fluctuations in between -- one can conclude that mass
formulas of the type Eq.~\eqref{eq:masses} do encode the role of the
nuclear deformation.  Shell effects are likely responsible for most of
these residuals, and we shall discuss how to include these toward the
end of this paper. A sufficiently accurate theory of nuclear masses
may even aim to include contributions arising from quantum chaos.

\begin{subequations}
  We will describe a \gls{NEDF}, which depends on the smallest number
  of phenomenological parameters to account for all the contributions
  in the nuclear mass formulas Eqs.~\eqref{eq:Bethe} and
  \eqref{eq:masses}, with the exception of the even-odd staggering due
  to pairing effects.  First we relate these parameters to various
  physical quantities relevant for nuclear physics.  For a large
  nucleus, the Coulomb energy can be used to estimate the saturation
  density $\rho_0$ by approximating the nucleus as a uniformly charged
  sphere with $E_C = 3Z^2e^2/5R =a_CZ^2/A^{1/3}$, where $R=r_0A^{1/3}$
  and $r_0\approx \SI{1.2}{fm}$ is a nuclear length scale:
  \begin{gather}
    \rho_0 =\frac{3}{4\pi r_0^3}, \quad \text{where} \quad r_0 = \frac{3e^2}{5a_C}.
  \end{gather}
  One can further estimate the ground-state energy of infinite nuclear matter per
  nucleon $\varepsilon_0$, the nuclear surface tension $\sigma$, and their
  dependence on the isospin $(N-Z)/2$:
  \begin{align}
    \varepsilon_0 &= \frac{E(N, Z)}{A} = a_v + a_I\frac{(N-Z)^2}{A^2}, \\
    \sigma &= a_s + a'_I\frac{(N-Z)^2}{A^2}.
  \end{align}
  Finally, one can relate the value of the coefficient $a'_C$ (or of the alternative
  coefficient of the contribution $a''_CZ^2/A$ to the mass formula~\cite{Myers:1969}) with the
  nuclear surface diffuseness.

  One thus expects a \gls{NEDF} as
  accurate as the mass formula to contain not more than 5 or 6
  significant parameters.  As we shall see, such a functional does
  exist, and performs comparably well with as few as 4 parameters however.
  That a functional dependent on such a small number of phenomenological
  parameters can go far beyond the capabilities of mass formulae
  to describe density distributions and dynamics is truly remarkable.
\end{subequations}

\subsection{Form of the NEDF}
We postulate a \gls{NEDF} with three main contributions, which
significantly improves on the Weizsäcker's original
idea~\cite{Weizsacker:1935}:
\begin{gather}
  \label{eq:NEDF}
  \mathcal{E}[\rho_n,\rho_p] =
  \mathcal{E}_{\text{kin}}[\rho_n,\rho_p] +
  \mathcal{E}_{C}[\rho_n,\rho_p] +
  \mathcal{E}_{\text{int}}[\rho_n,\rho_p].
\end{gather}
As we shall now discuss, the first two terms are well motivated by
a semi-classical expansion and electrodynamics and they have no free parameters.
All of the phenomenological parameters occur in the interaction piece alone.

\paragraph*{Kinetic Terms:}
The kinetic energy density follows from a semi-classical expansion of the
non-interacting system~\cite{Brack:1997,Dreizler:1990lr}, and is expressed
in terms of the neutron/proton number
densities $\rho_{n,p}$ and masses $m_{n,p}$:
\begin{multline}
  \label{eq:NEDF_kin}
  \mathcal{E}_{\text{kin}}[\rho_n,\rho_p]
  = \\
  \sum_{\tau=n,p} \frac{\hbar^2}{2m_\tau}
  \left[
    \frac{1}{9} \Abs{\grad{\rho^{1/2}_\tau}}^2
    + \tfrac{3}{5}(3\pi^2)^{2/3} \rho_\tau^{5/3} \right] + \cdots.
\end{multline}
Higher order corrections to this are considered in
Eq.~\eqref{eq:grad4} below.  The factor of $1/9$ can be derived for
smoothly varying densities, and should be compared with the factor of
unity originally suggested by \textcite{Weizsacker:1935}.

\paragraph*{Coulomb Terms:}
\begin{subequations}
  \label{eq:NEDF_C}
  The direct Coulomb energy and exchange contribution in the Slater
  approximation are:
  \begin{align}
    \mathcal{E}_{C}[\rho_n, \rho_p] &=
    \frac{1}{2} V_C(\vect{r})\rho_{\text{ch}}(\vect{r})  - \frac{e^2\pi}{4} \left(
      \frac{3\rho_p(\vect{r})}{\pi}\right)^{4/3},
    \label{eq:Slater}\\
    V_C(\vect{r}) &= \int \d^3r' \frac{\rho_{\text{ch}}(\vect{r}')}
                                      {\abs{\vect{r} - \vect{r}'}},
  \end{align}
  where $e$ is the proton charge, and $\rho_{\text{ch}}$ is the charge
  density, which is obtained from the proton and neutron densities by
  convolving with the appropriate charge form factors:
  \begin{gather}
    \label{eq:FF}
    \rho_{\text{ch}} = G^n_E * \rho_n + G^p_E * \rho_p.
  \end{gather}
  The charge form factors are determined experimentally, and we
  approximate the Fourier transforms of the form factors with the
  dipole term for the proton, ${G}^{p}_E(Q) \approx (1 +
  Q^2/\SI{0.71}{GeV^2})^{-2}$~\cite{Perdrisat:2007}, and
  ${G}_E^{n}(Q) \approx a(1 + Q^2r_+^2/12)^{-2} - a(1 +
  Q^2r_-^2/12)^{-2}$ with $r_\pm^2 = r_{\text{avg}}^2 \pm
  \braket{r_n^2}/2a$, $\braket{r_n^2} = -\SI{0.1147(35)}{fm^2}$,
  $r_{\text{avg}} = \SI{0.856(32)}{fm}$, and
  $a =\num{0.115(20)}$~\cite{Gentile:2011}.  Including these form factors does not
  significantly improve the mass fits, but improves somewhat the fit
  of the charge radii.  In principle, one might allow the coefficient
  of the Coulomb exchange term to vary; this is done, for example, in
  atomic physics to obtain better estimates of the Coulomb exchange
  energy.  We find, however, that fitting the nuclear binding energies
  leads with high accuracy to the same coefficient presented in
  Eq.~\eqref{eq:Slater}, so we leave it fixed and do not include this
  as a parameter in our model.
\end{subequations}

We require our energy density functional to be an isoscalar and include no isospin
breaking terms other than those due to the neutron-proton mass difference and
the Coulomb interaction. Additional isospin violation due to
up and down quark mass differences and electromagnetic
effects~\cite{Miller:1990iz, Miller:2006tv} beyond these two contributions are
much smaller and are partly responsible mainly for the Nolen-Schiffer
anomaly~\cite{Nolen:1969}, to which the screening of the Coulomb exchange
also contributes at a comparable level~\cite{Bulgac:1996}.

\paragraph*{Interaction Terms:}
\begin{subequations}
  \label{eq:NEDF_int}
  All the undetermined parameters of the model
  appear in the interaction terms.  We parameterize these as
  \begin{gather}
    \begin{multlined}
      \mathcal{E}_{\text{int}}(\rho_n, \rho_p) =
      (\eta - \tfrac{1}{9})
      \sum_{\tau=n,p}\frac{\hbar^2}{2m_\tau}\Abs{\grad\rho^{1/2}_\tau}^2\\
      + \sum_{j=0}^2\mathcal{E}_j(\rho)\beta^{2j}
      + \mathcal{E}_{\text{entrain}}(\rho_n,\rho_p),
    \end{multlined}\\
    \mathcal{E}_j(\rho) =  a_j \rho^{5/3} + b_j \rho^2 + c_j \rho^{7/3},
    \label{eq:abc}
  \end{gather}
  where $\rho$ is the total density, and $\beta$ is the asymmetry:
  \begin{gather}
    \rho = \rho_n + \rho_p, \qquad
    \beta = \frac{\rho_n - \rho_p}{\rho_n + \rho_p}.
  \end{gather}
  We include a correction to the gradient term in
  $\mathcal{E}_{\text{kin}}$ so that the overall coefficient of the gradient term is
  $\eta$, and up to 9 parameters $a_j$, $b_j$, and $c_j$ for $j\in\{0,
  1, 2\}$ describing the equation of state for homogeneous nuclear
  matter.  Three of these (for $j=2$) will be fixed by the equation of
  state of neutron matter determined in ab initio calculations (see
  section~\ref{sec:infin-nucl-neutr}). Three of the remaining six
  parameters ($a_0$, $c_1$, and a combination of $a_1$ and $b_1$) are
  found to be only marginally significant at the level of changing the energy rms by $\delta
  \chi_E < \SI{0.1}{MeV}$, so that in the end we shall be left with only 4 significant
  parameters: $\eta$, $b_0$, $c_0$, and a combination of $a_1$ and
  $b_1$.

  We shall later include an entrainment term
  $\mathcal{E}_{\text{entrain}}$ with a single parameter $\alpha$, the
  form and meaning of which will be elucidated in
  section~\ref{sec:dynamical-properties}, see
  Eq.~\eqref{eq:entrainment}.  We find this term to have very little
  significance for the static properties of nuclei, but we will show
  that it has a pronounced effect on isovector dynamics, such as the
  \gls{GDR} mode.  For the moment while we discuss static properties
  we set it to zero $\alpha = 0$. One more parameter, $\xi$, will be discussed
  when we will discuss the computation of shell effects in section~\ref{sec:shell-effects}.
\end{subequations}

\begin{subequations}
  The equations that determine the equilibrium densities of a nucleus are
  obtained by minimizing the energy of a given nucleus
  $E(N,Z) = \int\d^3r \mathcal{E}[\rho_n,\rho_p] $ with respect to the densities, while
  constraining the total numbers of neutrons $N$ and protons $Z$  with two chemical
  potentials $\mu_{n,p}$:
  \begin{gather}
    -\eta\frac{\hbar^2}{2m_\tau} \laplacian\rho^{1/2}_\tau + U_\tau\rho^{1/2}_\tau =
    \mu_\tau\rho^{1/2}_\tau, \label{eq:psi}\\
    U_\tau=\frac{ \partial \mathcal{E} [\rho_n,\rho_p] }{ \partial \rho_\tau},
    \text{ for } \tau \in \{n,p\}.
  \end{gather}
\end{subequations}

\subsection{Gradient Terms}
\textcite{Weizsacker:1935} originally introduced the term proportional
to gradients of the densities $\abs{\vect{\nabla}\rho_\tau^{1/2}}^2$ with a
value of $\eta = 1$ that was later shown to be valid only if the density has small
amplitude rapid oscillations~\cite{Brack:1997,Dreizler:1990lr,Jones:1989}.  It was later rigorously
proven that a semi-classical expansion of the non-interacting fermions
in the extended Thomas-Fermi approximation (the limit of a slowly
varying external potential) yields the value $\eta_{\text{TF}} = 1/9$,
which defines the lowest order gradient contributions in
$\mathcal{E}_{\text{kin}}$~\eqref{eq:NEDF_kin}~\cite{Dreizler:1990lr,
  Brack:1997, Jones:1989}.  We treat the coefficient $\eta$ as a phenomenological
parameter since gradient terms can also be generated by
interactions~\cite{Negele:1972, Negele:1975, Gebremariam:2010x}.
Fitting the nuclear masses yields values of $\eta$ close to \num{0.5},
roughly half-way between the semi-classical and Weizsäcker values.

A semi-classical expansion of non-interacting
fermions~\cite{Dreizler:1990lr, Brack:1997} yields the following
higher order corrections to $\mathcal{E}_{\text{kin}}$, that we have
not included in Eq.~\eqref{eq:NEDF_kin}:
\begin{gather}
  \mathcal{E}_{\grad^4}(\rho_n, \rho_p) = \sum_{\tau\in\{n,p\}}
  \frac{\hbar^2}{2m_\tau}\frac{1}{810(3\pi^{2})^{2/3}}f(\rho_\tau),
  \label{eq:grad4}\\
  f(\rho) = \rho^{1/3}\left[
      \left(
        \frac{\grad{\rho}}{\rho}
      \right)^4
      - \frac{27}{8}\left(
        \frac{\grad{\rho}}{\rho}
      \right)^2
      \frac{\grad^2 \rho}{\rho}
      + 3\left(
        \frac{\grad^2 \rho}{\rho}
      \right)^2
    \right]. \nonumber
\end{gather}
This type of correction has been studied in nuclear physics and shown
to lead to quite accurate estimates of the kinetic energy density
within the extended Thomas-Fermi approximation~\cite{Brack:1976,
  Brack:1985,Brack:1997}.  As within a \gls{DFT}, such terms can also
arise due to the finite range of the interactions in a matter similar
to some Skyrme interactions~\cite{Negele:1972, Negele:1975,
  Gebremariam:2010x}.  However, these terms -- even with adjustable
parameters -- do not significantly change the quality of the mass
fits, so we do not consider them in our main analysis.  Including them
perturbatively in the fit, however, does improve the fit of the charge radii. For
example, fitting the overall coefficient reduce the charge radii
residual $\chi_r$ (see details in Section~\ref{sec:fitt-mass-charge})
from $\chi_r\approx \SI{0.14}{fm}$ to $\chi_r\approx \SI{0.09}{fm}$.
Fitting each of the three terms independently further reduces the
residuals to $\chi_r\approx \SI{0.06}{fm}$. We do not include such
fourth-order terms in our functional as they can lead to a complex
behavior of the emerging equation for the densities,
which can be difficult to rationalize. (See, for example, the
analysis of fourth order differential equations arising in case of non-local potentials by
\textcite{Bulgac:1988x}.)  Higher order gradient corrections than
Eq.~\eqref{eq:grad4} lead to an unphysical behavior of the densities
in the classically forbidden regions.  The
semi-classical expansion has an asymptotic
character~\cite{Jones:1989}. It is known
that corrections beyond second order do not always lead to
improvements~\cite{Jones:1989}.

We point out one more property that will be discussed in more detail
in section~\ref{sec:dynamical-properties}: a value of $\eta = 1/4$
corresponds to a dynamical theory of superfluid neutron and proton
pairs. One might naïvely have thought of $\eta \approx 0.5$ as an
effective nucleon pair mass $m_{\text{eff}} \approx 2 m$, but this
leaves the potential $U_\tau$ wrong by a factor of 2.  The parameter
$\eta$ may simply be thought of as a way to control the falloff of
the densities in the surface region where the interaction effects are
still strong.  One should not expect to obtain the correct asymptotic
behavior of the densities far from the nucleus, where the interactions
become vanishingly small. The main reason is that DFT is by definition
an integral approach, which minimizes globally the energy and not
particular local quantities.

A gradient term alone of the form $\abs{\grad{\rho}}^2$ instead of
$\abs{\grad{\rho^{1/2}}}^2 = \abs{\grad{\rho}}^2/4\rho$ leads to
unphysical density profiles, with a discontinuity in $\grad{\rho}$
at a finite radius, beyond which the density vanishes exactly.

\begin{figure}[htb]
  \includegraphics[width=0.9\columnwidth]{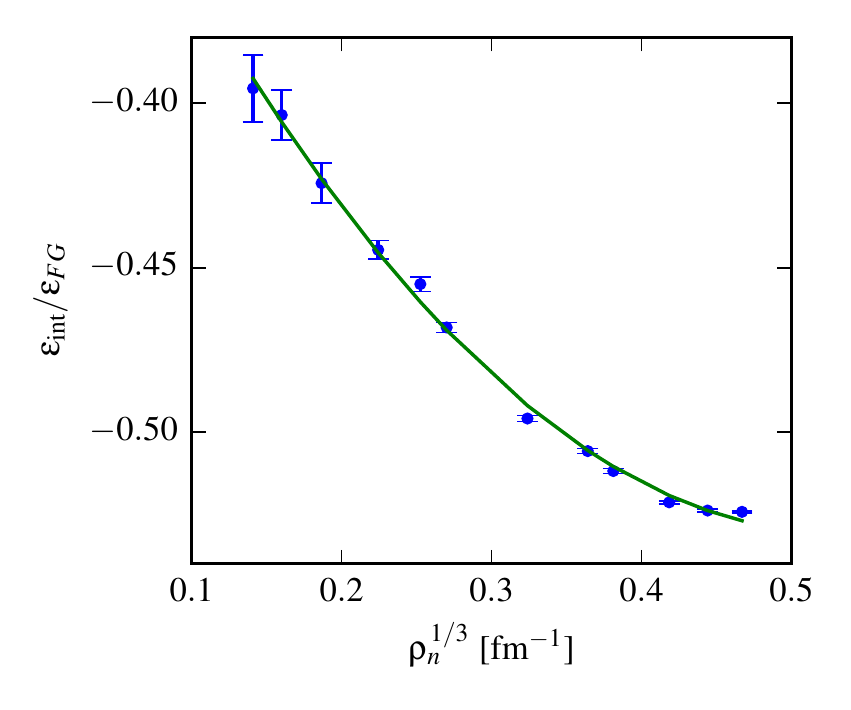}
  \caption{\label{fig:QMC}%
    (Color online) The \gls{QMC} results of \textcite{Wlazowski:2014a}
    for the interaction energy per neutron displayed as the ratio
    $\epsilon_n/\epsilon_{FG}$ defined in Eq.~\eqref{eq:nm} (with
    $\beta = 1$), where $\epsilon_{FG} = 3 \hbar^2
    (3\pi^2\rho_n)^{2/3}/(10m_n)$.  If $a_n=0$ in
    Eq.~\eqref{eq:nm}, the ratio $\epsilon_{\text{int}}/\epsilon_{FG}$ would tend to 0 for
    $\rho_n\rightarrow 0$. For densities
    $\rho_n^{1/3}\abs{a_{NN}} < 1$
    (where $a_{NN}=\SI{18.9}{fm}$ is the $s$-wave neuttron-neutron
    scattering length)
    the leading order correction to the kinetic
    energy density per particle contribution would be  instead
    linear in density  $4\pi\hbar^2 a_{NN}\rho_n/m_n$ .}
\end{figure}

\subsection{Infinite Nuclear and Neutron Matter}%
\label{sec:infin-nucl-neutr}
In infinite homogeneous nuclear matter, as might be found in a neutron star for
example, the gradient and Coulomb terms vanish (charge neutrality is maintained
by a background of electrons).  Neglecting the small neutron-proton mass
difference $m_n \approx m_p = m$, the functional acquires a simpler form:
\begin{multline} \label{eq:inm}
  \mathcal{E}(\rho_n, \rho_p) =
  \frac{3\hbar^2(3\pi^2)^{2/3}}{10m}(\rho_n^{5/3} + \rho_p^{5/3})\\
  +
  \sum_{j=0}^{2} \left(a_j\rho^{5/3} +  b_j\rho^2 + c_j
    \rho^{7/3}\right)\beta^{2j},
\end{multline}
This portion of the functional is essentially an expansion in powers of the
Fermi momenta $k_F$: $k_{n,p} = (3\pi^2 \rho_{n,p})^{1/3}$ with three terms
only $k_F^{5}, k_F^6$ and $k_F^{7}$,
an ubiquitous power expansion in many-body perturbation theory.
It is also supported by fitting the neutron matter  equation of state
($\rho_p=0$, $\beta=1$):
\begin{subequations}
\begin{align}
 \varepsilon_n(\rho_n)&=
  \frac{3\hbar^2}{10m_n} (3\pi^2\rho_n)^{2/3}+\varepsilon_\text{int}(\rho_n), \label{eq:NM}\\
   \varepsilon_\text{int}(\rho_n) &= a_n\rho_n^{2/3}
  +  b_n\rho_n + c_n  \rho_n^{4/3},\label{eq:nm}
 \end{align}
 where
\begin{gather}
  \begin{alignat}{3}
    a_n &= a_0 + a_1 + a_2 &= && \SI{-32.6}{MeV fm^2}&, \\
    b_n &= b_0 + b_1 + b_2 &= && \SI{-115.4}{MeV fm^3}&, \\
    c_n &= c_0 + c_1 + c_2 &= && \SI{109.1}{MeV fm^4}&.
  \end{alignat}
\end{gather}
\end{subequations}
These values have been obtained by fitting the neutron matter equation
of state as calculated with \gls{QMC} including up to N$^3$LO two-body
and up to N$^2$LO three-body interactions from chiral perturbation
theory~\cite{Wlazowski:2014a}, as shown in Fig.~\ref{fig:QMC}.
Thus, we can include directly the fixed $j=2$ parameters $a_2$, $b_2$,
and $c_2$ from the values of $a_n$, $b_n$, and $c_n$ describing the
\gls{QMC} results without adding additional free parameters to the
\gls{NEDF}.

As we shall discuss in section~\ref{sec:fitt-mass-charge}, adding the
quartic $\beta^4$ $(j=2)$ terms does not significantly affect the quality of the
mass fits since most nuclei realize small asymmetries $\beta < 0.25$.
This demonstrates an important point: nuclear masses do not constrain
the quartic terms  $\beta^4$.  Higher powers thus provide a
direct (and independent) handle on the equation of state of neutron
matter.

At this time we do not have an equally accurate \gls{QMC} calculation
of nuclear matter with varying isospin composition, so we must rely
instead on a phenomenological approach.  Our main assumption is that
we can describe both the isoscalar ($j=0$) and isovector ($j=1$, $\beta^2$) parts
of the nuclear equation of state in a similar fashion to the equation
of state of pure neutron matter, using the same three powers of the
corresponding Fermi momenta. This approach differs from typical
Skyrme-like parameterizations, which include terms with higher powers
of densities (e.g. $\rho^{8/3}$ arising from $\tau\rho$ type of terms,
where $\tau$ is kinetic energy density).

The terms $a_j\rho^{5/3}$ are somewhat unexpected and they are not
included in Skyrme-like parameterizations.  \textcite{Tondeur:1978}
introduced only a term $a_1$ (without theoretical justification), but
it makes sense to include the other $a_j$ for several reasons.  The
\gls{QMC} calculations of \textcite{Gezerlis:2010a, Wlazowski:2014a,
  Gandolfi:2015}, see Fig.~\ref{fig:QMC}, are consistent with the
  existence of a non-vanishing
parameter $a_n$ in the neutron equation of state, which implies that
$a_n=\sum_{j=0}^2 a_j\neq 0$.  They also appear naturally in the case
of the unitary Fermi gas~\cite{Zwerger:2011}, a fact confirmed to high
precision in many experiments.  The unitary Fermi gas is a system of
two species of fermions, interacting with an $s$-wave interaction with
zero range and infinite scattering length. In response to the
Many-Body X challenge posed by Bertsch in 1999, \textcite{Baker:1999}
showed that the system was stable.  The energy density of the unitary
Fermi gas scales exactly like the kinetic energy density of a free
Fermi gas $\mathcal{E} \propto\rho^{5/3}$.  Since both neutron and
protons have similar $s$-wave interaction properties, one expects the
nuclear energy density to behave somewhat like the unitary Fermi
gas.

Although the energy density of the unitary Fermi gas scales as the
kinetic energy, this is not necessarily due to a mass renormalization
as one might naïvely suspect and the answer is not a clear cut one
unless additional information is obtained.
The \gls{QMC} calculations of the
single quasi-particle dispersion~\cite{Carlson:2005kg} and the
calculation of the spectral
weight function~\cite{Magierski:2009, Magierski:2011}, both find
almost the same effective mass $\approx m$ in the unitary Fermi gas,
very close to unity.
However, this does not preclude the interpretation of some part of this
term as arising from the kinetic energy density $\tau$ as is the case
in the unitary Femi gas too~\cite{Bulgac:2007a, Bulgac:2011, Bulgac:2013b}.
The QMC calculations  are not of enough accuracy to either completely
exclude or confirm an effective mass different from unity.

Furthermore, \gls{QMC} calculations of dilute neutron matter at both
zero and finite temperatures point to a very large pairing gap $\Delta
\approx 0.25\epsilon_F$~\cite{Gezerlis:2010a} and $\Delta = 2.8
T_c$~\cite{Wlazlowski:2010, Wlazlowski:2011,Abe:2009a,Abe:2009b},
where $\epsilon_F$ and $T_c$ are the Fermi energy and critical
temperature for the superfluid to normal transition.  In comparison,
\gls{BCS} superfluids or mean-field approaches lead $\Delta \approx
1.76 T_c$, indicating that neutron matter is not a simple \gls{BCS}
superfluid~\cite{Magierski:2009, Magierski:2011}. This is further
supported by evidence for a pseudogap at temperatures above $T_c$
where Cooper pairs and a significant depletion of the density of
states still exist even in the absence of long-range order.  The
properties of neutron matter are thus closely related to those of high
$T_c$ superconductors.  The common approach of adding pairing
correlations within an \gls{HFB} or \gls{BCS} approximation must thus
be considered only qualitatively valid, and properly tuned \gls{DFT}
approach is required for quantitative results~\cite{BY:2002fk,
  Bulgac:2002uq, BY:2003, Yu:2003, Yu:2003b, Bulgac:2007a,
  Bulgac:2009, Bulgac:2011c, Bulgac:2011b, Bulgac:2011, Stetcu:2011,
  Stetcu:2014, Bulgac:2014x, Wlazlowski:2015}.

\begin{table*}[pt]
  \begin{ruledtabular}
    \begin{tabular}{cccccccccccccc}
      NEDF & {$\eta$}& {$\tilde{a}_{0}$}& {$\tilde{b}_{0}$}& {$\tilde{c}_{0}$}& {$\tilde{a}_{1}$}& {$\tilde{b}_{1}$}& {$\tilde{c}_{1}$}& {$\tilde{a}_{2}$}& {$\tilde{b}_{2}$}& {$\tilde{c}_{2}$}& {$\delta$}& {$\chi_{E}$}& {$\chi_{r}$} \\
           & & & & & & & & & & & & [\si{MeV}]& {[\si{fm}]} \\
      \hline
0& 0.4719& 0& -3.15166& 2.12378& 1.048& -0.610& 0& 0& 0& 0& 0.3246& 2.59& 0.145\\
1& 0.4742& 0& -3.13778& 2.10995& 0.981& -0.544& 0& 0& 0& 0& 0.3250& 2.58& 0.135\\
2& 0.4743& 0& -3.14595& 2.11873& 0.961& -0.521& 0& 0& 0& 0& 0& 2.71& 0.140\\
1r& 0.4807& 0& -2.98351& 1.94501& 1.087& -0.668& 0& 0& 0& 0& 0.3330& 2.71& 0.051\\
3& 0.4800& -0.088& -2.95408& 2.01459& 1.000& -0.549& -0.017& 0& 0& 0& 0.3234& 2.58& 0.139\\
3n& 0.4739& -0.061& -3.00477& 2.03797& 1.565& -1.367& 0.181& -1.765& 3.8811& -1.9731& 0.3279& 2.57& 0.133\\
3nr& 0.4815& -0.061& -2.86708& 1.89073& 1.563& -1.347& 0.169& -1.763& 3.7239& -1.8138& 0.3529& 2.67& 0.050\\
E& 0.4885& 0& -3.14903& 2.11957& 0.277& 0.277& 0& 0& 0& 0& 0.3177& 2.64& 0.129\\
Er& 0.4957& 0& -3.00642& 1.96664& 0.264& 0.264& 0& 0& 0& 0& 0.3600& 2.74& 0.051\\[1.5ex]
En& 0.4866& 0& -3.15157& 2.12271& 0.272& 0.272& 0& -0.533& 2.3889& -1.8763& 0.3192& 2.62& 0.133\\
Enr& 0.4970& 0& -3.00488& 1.96492& 0.260& 0.260& 0& -0.521& 2.2540& -1.7185& 0.3545& 2.74& 0.051
    \end{tabular}
  \end{ruledtabular}
  \caption{\label{table:NEDF_A} Dimensionless fit parameters and residuals for the various NEDFs.
      Parameters have been scaled by appropriate powers of
      $\rho_0 = 0.15\si{fm^{-3}}$ and
      $\epsilon_F = \tfrac{\hbar^2}{2m} (3\pi^2\rho_0/2)^{2/3} = 
      \SI[round-precision=5]{35.2941961307}{MeV}$:
      $\tilde{a}_j = a_j\rho_0^{2/3}/\epsilon_F$,
      $\tilde{b}_j = b_j\rho_0/\epsilon_F$, 
      and $\tilde{c}_j = c_j\rho_0^{4/3}/\epsilon_F$.}
\end{table*}

\begin{figure}[tb]
  \includegraphics[width=0.81\columnwidth]{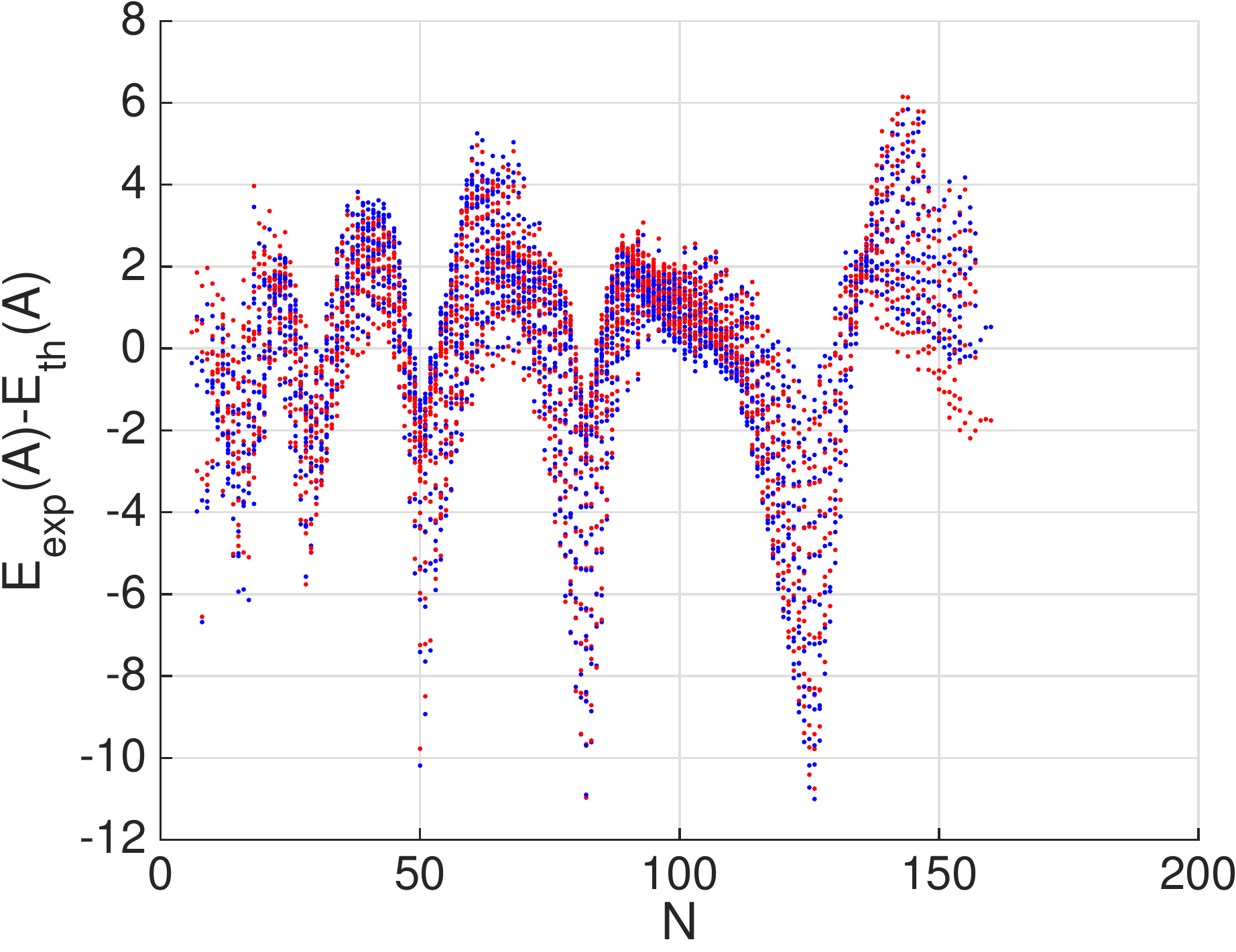}
  \includegraphics[width=0.81\columnwidth]{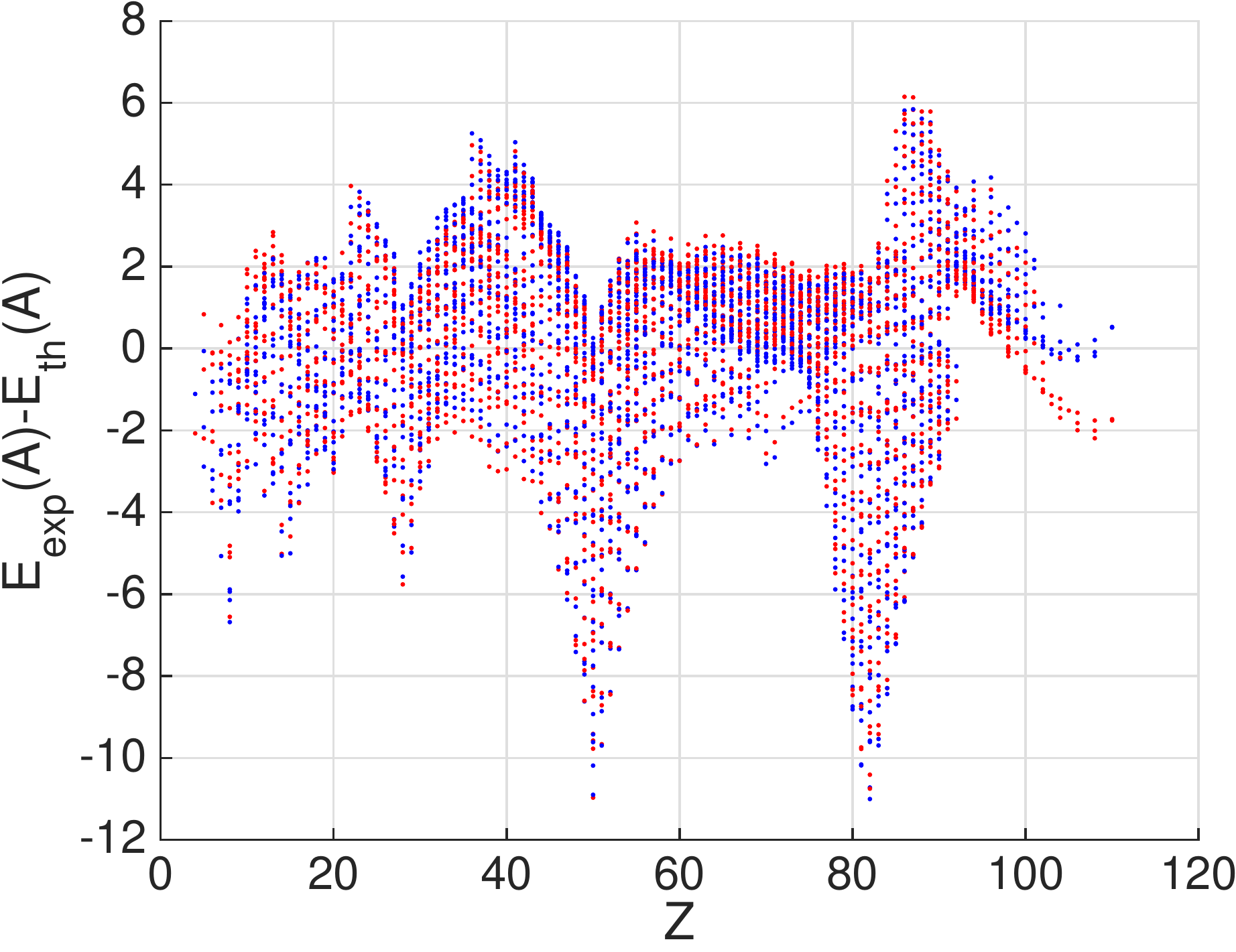}
  \includegraphics[width=0.81\columnwidth]{bezn10}
  \caption{\label{fig:be_zn}%
    (Color online) The blue dots show the results obtained using fit NEDF-1
    with $\chi_E = \SI{2.58}{MeV}$, while the red dots are the results of the
    fits using nuclear mass formula Eq.~\eqref{eq:masses}, with $\chi_E =
    \SI{2.52}{MeV}$. When compared against each other, the \gls{rms} energy
    deviation between the two fits is $\chi_E = \SI{0.36}{MeV}$.  Thus,
    \gls{NEDF}-1 essentially reproduces the nuclear mass formula
    Eq.~\eqref{eq:masses}. The lower panel is a rendering of calculated ground
    state energy differences between the experimental measured values and NEDF-1
    results, in which one can see clearly the magic
    numbers separately for neutrons and protons. }
\end{figure}

\section{Fitting Masses and Charge Radii}%
\label{sec:fitt-mass-charge}
We fit our \glspl{NEDF} to the $N_E = 2375$ measured (not
extrapolated) nuclear masses with $A\geq 16$ from~\cite{Audi:2012,
  Wang:2012}.  We also consider the $N_r = 883$ matching charge
nuclear radii from~\cite{Angeli:2013} with $\chi_r^2 = \sum\abs{\delta
  r}^2/N_r$.  When we include the charge radii in the fit, we minimize
the following quantity $\chi^2_E/(\SI{3}{MeV})^2 +
\chi^2_r/(\SI{0.05}{fm})^2$ which roughly equalizes the weight of the
mass and radii contributions in the fit.

At this point, we have 7 parameters in our \gls{NEDF}: $\eta$,
$a_{0,1}$, $b_{0,1}$, and $c_{0,1}$ (the $j=2$ parameters are fixed by
the neutron matter equation of state).  In addition, we include by
hand the conventional even-odd staggering Eq.~\eqref{eq:odd-even} with
a coefficient $\delta$ to describe pairing correlations, even though
this has very little significance in the fits.  We consider the
following fits:
\begin{description}
\item[NEDF-0] A six parameter least-squares fit of the $N_E = 2375$ nuclear
  masses~\cite{Audi:2012, Wang:2012} including $\eta$, $b_0$, $c_0$, $a_1$,
  $b_1$, and $\delta$ but setting the nucleon charge form factors Eq.~\eqref{eq:FF}
  ${G}^{p}_{E} \equiv 1$ and ${G}^{n}_{E} \equiv 0$.

\item[NEDF-1] The same as NEDF-0, but including the measured charge
  form factors.  Comparing with NEDF-0 we see that the electric form factors
  are not significant for the overall mass fits, but slightly impact the charge
  radii at the \SI{0.01}{fm} level (for the reduced $\chi_r$).

\item[NEDF-2] The same as NEDF-1, but without the pairing parameter $\delta=0$.
  Comparing with NEDF-1 we see that odd-event staggering is also relatively
  insignificant at the level of \SI{0.1}{MeV} per nucleus.  This is consistent
  with the results from the mass formulas in Table~\ref{table:liquid_drop}.

\item[NEDF-1r] The same as NEDF-1, but including the $N_r = 883$ charge radii
  into the fit. We see that there is significant room to improve the
  description of the charge radii without significantly degrading the mass
  fits.  The charge radii residuals for NEDF-1 and NEDF-1r are plotted in
  Fig.~\ref{fig:chrad}.

\item[NEDF-3] The same as NEDF-1, but with all 8 parameters, including
  $a_0$ and $c_1$ that we omitted from the previous fits.  In
  conjunction with the principal component analysis shown in
  Fig.~\ref{fig:PCA}, this fit demonstrates that the terms with
  parameters $a_0$ and $c_1$ are insignificant.

\item[NEDF-3n] The same as NEDF-1, but with all 8 parameters, including
  $a_0$ and $c_1$ that we omitted from the previous fits, and the
  $\beta^4$ parameters for the terms quartic in isospin,
  constrained by the \gls{QMC} neutron matter
  equation of state~\cite{Wlazowski:2014a} using Eqs.~\eqref{eq:nm}.
  That the quality of the fit, isoscalar, and isovector parameters
  change very little, demonstrates that the neutron matter equation of
  state is essentially independent of the nuclear masses.
\item[NEDF-3nr] The same as NEDF-3n but including the charge radii as in fit
  NEDF-1r.  That the $a_0$ and $c_1$ terms are insignificant for both masses and
  radii is also emphasized by this fit.
\end{description}
Finally, we consider two parameter sets that represent the minimal
functionals. Each has 4 significant parameters (and $\delta$, which is
insignificant):
\begin{description}
\item[NEDF-E] Following the principal component analysis of NEDF-3n
  (discussed below) we find the combination $a_1 - b_1\rho_*^{1/3}$ to
  be only weakly constrained by the mass fit.  To test this, we set
  $a_1 = b_1\rho_*^{1/3}$ where $\rho_* = \SI{0.15}{fm^{-3}}$ is a
  constant.  The combination $a_1 - b_1\rho_*^{1/3}$, to which the
  masses are insensitive, allows independent control the slope $L_2$ of
  the symmetry energy (see Eq.~\eqref{eq:L}).  From the fits we see
  that this same combination also controls the neutron skin
  thicknesses.
\item[NEDF-Er] The same as NEDF-E but including the charge radii as in
  fit NEDF-1r.
\item[NEDF-En] This is our main fit.  It is the same as NEDF-E but
  includes the $\beta^4$ parameters adjusted to reproduce the neutron
  matter equation of state as in fit NEDF-3n.
\item[NEDF-Enr] The same as NEDF-En but including the charge radii as in
  fit NEDF-1r.
\end{description}
These fits are summarized in Table~\ref{table:NEDF_A} with the saturation and
symmetry properties in Table~\ref{table:NEDF_B}.  The residuals for fit
NEDF-1 are shown in Fig.~\ref{fig:be_zn} and compared with a fit to the nuclear
with mass formula Eq.~\eqref{eq:masses}.

The reduced $\chi_E$ for these fits is comparable to that obtained
using the nuclear mass formulas Eq.~\eqref{eq:Bethe} with 4 parameters
(plus $\delta$)
and Eq.~\eqref{eq:masses} with 5 parameters (plus $\delta$). This is
consistent with our hypothesis that a \gls{NEDF} for masses should
contain no more than 5 significant parameters.
Note, however, that unlike the mass formulas, the \gls{NEDF}
also gives a good description of charge radii -- for which the mass
formula says nothing -- and provides access to nuclear dynamics as we
shall discuss in section~\ref{sec:dynamical-properties}.

\subsection{Discussion}
The accuracy of the ground state binding energies obtained using this
\gls{NEDF}, see Fig.~\ref{fig:be_zn} and Table~\ref{table:NEDF_A}, compares
extremely well with much more sophisticated self-consistent approaches that
naturally account for the shell-corrections.

The UNEDF2 nuclear energy functional introduced by
\textcite{Kortelainen:2014, Kortelainen:2014x} has a residual of
$\chi_E = \SI{1.95}{MeV}$ per nucleus for \num{555} even-even
nuclei. Since the UNEDF2 functional is formulated in terms of
single-nucleon orbitals, one would expect it to account for
shell-corrections, but this functional still displays large
discrepancies of the binding energies of magic nuclei (close to
\SI{8}{MeV} in case of \ce{^{208}Pb}).  The UNEDF2 functional depends on
11 parameters, with several additional parameters to describe pairing
correlations. The UNEDF2 functional also leads to a larger pairing coupling
constant for protons than for neutrons, thus violating isospin
invariance and the expectation that Coulomb effects reduce proton
pairing, an expectation which appears to agree with
experiments~\cite{Lesinski:2009}.  A natural solution might be to
include in the \gls{NEDF} a pairing contribution of the form
\begin{subequations}
  \begin{gather}
    \begin{multlined}
      \mathcal{E}_{\text{pairing}} =
      g_0(\rho_n,\rho_p)\left [ \abs{\nu_n}^2+\abs{\nu_p}^2\right ] \\
      + g_1(\rho_n,\rho_p)\left [ \abs{\nu_n}^2-\abs{\nu_p}^2 \right ]
      [\rho_n-\rho_p],
    \end{multlined} \\
    g_{0,1}(\rho_n,\rho_p)=g_{0,1}(\rho_p,\rho_n),
  \end{gather}
\end{subequations}
where $\nu_{n,p}$ are the anomalous neutron and proton densities
respectively.  Since in measured nuclei one has predominantly $N\ge
Z$, see Fig.~\ref{fig:driplines}, a phenomenological analysis that
leads to an apparent coupling for protons larger than the one for
neutrons can be reconciled with coupling constants $g_0(\rho_n,\rho_p)
< 0$ and $g_1(\rho_n,\rho_p) >0 $.

\textcite{Baldo:2008x, Baldo:2013x} introduced an energy density
functional based on information extracted from \gls{QMC} calculations
of neutron and symmetric nuclear matter, and a few additional
parameters to describe pairing and spin-orbit interaction and finite
range effects; they assume that no quartic terms in isospin $\beta^4$ are
present in the \gls{NEDF}.  Depending on how they fit various subsets
of about 580 nuclei, they find $\chi_E$ ranging from
\SIrange{1.3}{2.4}{MeV}.

\textcite{Goriely:2009, Goriely:2013} have produced over the years a
series of \gls{NEDF} in which they obtain an even lower $\chi_E$
throughout the entire mass table as low as \SI{0.5}{MeV}.  However
this was obtained by adding a number of phenomenological corrections,
a procedure which so far has not been adopted by the other
practitioners using microscopic approaches.

Within  \gls{RMFT}  of nuclei in one of the best
parametrizations of the \gls{NEDF} for nuclear
masses~\cite{Niksic:2011, Agbemava:2014} one achieves a $\chi_E$
between 2 and 3 MeV for even nuclei using the AME2012 data
set~\cite{Audi:2012,Wang:2012}.

\subsection{Neutron Matter}
The inclusion of the $j=2$ terms quartic in $\beta^4$ in fits NEDF-3n
and NEDF-3rn demonstrates an important point: the equation of state of
neutron matter has very little impact on the form of the \gls{NEDF} at
the level of quadratic isovector contributions (see also the previous
section).  We have attempted to perform a fit of the nuclear binding
energies by including the quadratic terms in $\beta$ only, defining
the coefficients of the functional through the neutron equation of
state ($a_1=a_n-a_0$, $b_1 = b_n-b_0$, and $c_1 = c_n - c_0$), and
allowing $\eta$, $a_0$, $b_0$, and $c_0$ to vary. The energy rms was
at best $\chi_E = \SI{4.36}{MeV}$.

\begin{figure}[htbp]
  \includegraphics[width=0.9\columnwidth]{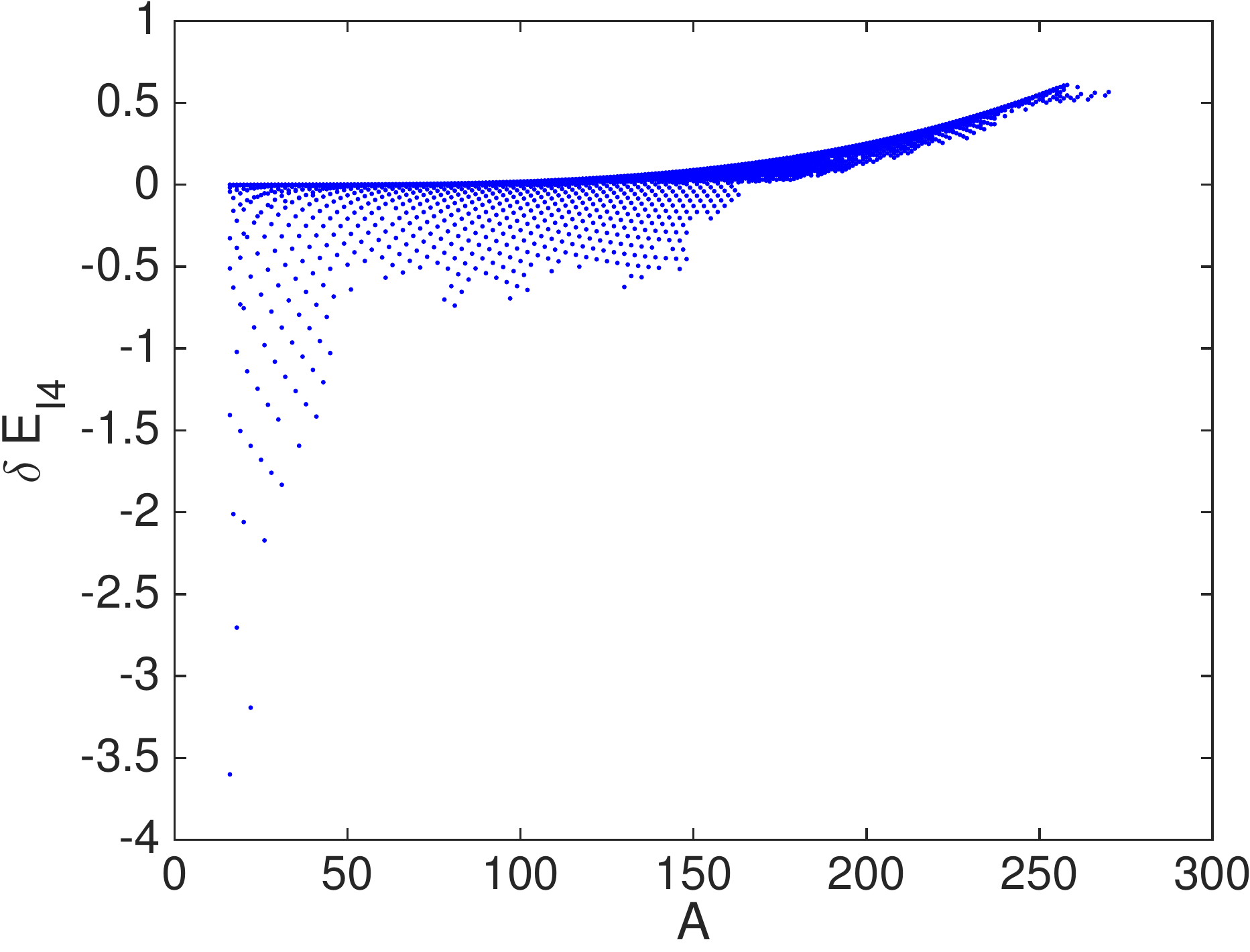}
  \includegraphics[width=0.9\columnwidth]{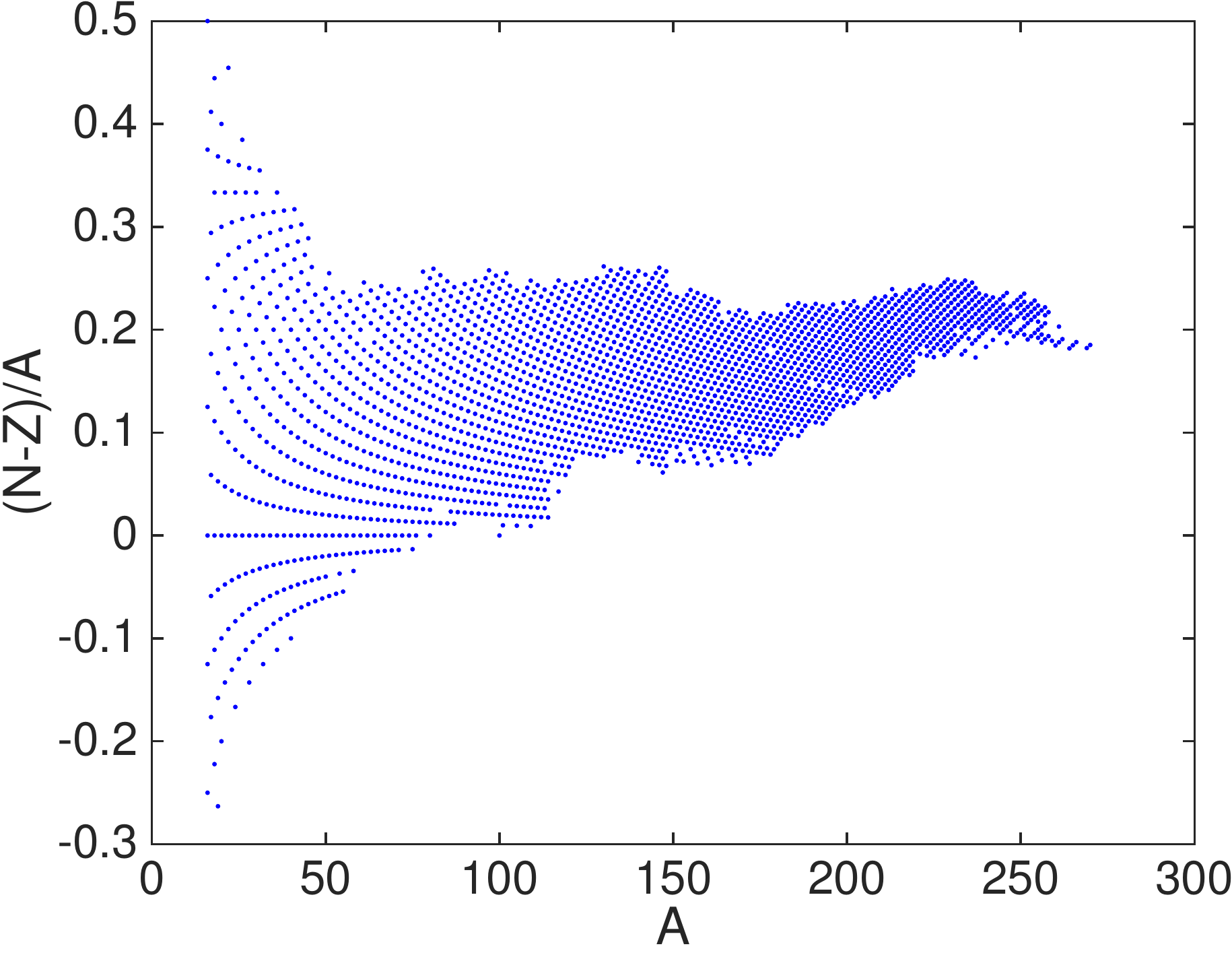}
  \caption{\label{fig:EI4}%
    (Color online) The contribution to the ground state
    energies of the terms quartic in isospin density $\delta E_{I4}=\int d^3
    \mathcal{E}_2(\rho)\beta^4$, evaluated perturbatively with NEDF-1,
    see Table~\ref{table:NEDF_A}. In the lower panel we display the
    ratio $(N-Z)/A$ for the nuclei we have considered.  Among the 2375
    nuclei we have considered, there are 33 nuclei with $N=Z$, 78
    nuclei with $Z>N$, and 70 nuclei with $\abs{N-Z}/A > 1/4$.
  }
\end{figure}

In measured nuclei, the ratio $\beta = (\rho_n-\rho_p)/\rho\approx (N-Z)/A$ is
$\abs{\beta} <0.25$ (with a very small number of exceptions), hence
nuclear masses are essentially insensitive to the presence of the
$\beta^4$ terms, as $\abs{\beta}^4 < 1/256$.  Only 46 nuclei have
$\abs{N-Z}/A > 1/4$ (mostly with $A<50$) and only 68 nuclei have $N\le
Z$ in the set we consider.  To assess the magnitude of these
effects, we have evaluated the $\beta^4$ contributions to the nuclear
binding energies perturbatively, see Fig.~\ref{fig:EI4}. This
contribution is quite small and can be easily overlooked when
discussing known nuclei, but is crucial in order to correctly
reproduce the energy of neutron matter.

Thus, using properties of the neutron matter to constrain the form of
the \gls{NEDF} and/or arguing against the inclusion of higher powers
of $(\rho_n-\rho_p)$~\cite{Fayans:1998, Baldo:2008x, Baldo:2013x,
  Brown:2013, Brown:2014, Goriely:2013, Fantina:2014, Reinhard:2010,
  Erler:2013} is an ill-advised procedure, and the applications of
functionals constructed in this manner, in particular to star
environments, should be regarded with suspicion.

\begin{table}[htbp]
  \sisetup{round-mode = places,
           round-precision = 3,
           } %
  \begin{ruledtabular}
    \begin{tabular}{ccccccccc}
           & &&&&&& \multicolumn{2}{c}{Neutron skin}\\
      NEDF & {$\rho_{0}$}& {$-\varepsilon_0$}& {$K$}& {$S$}& {$L$}& {$L_{2}$}& {\ce{^{208}Pb}}& {\ce{^{48}Ca}} \\
           & [\si{fm^{-3}}]& & & & & & [\si{fm}]& [\si{fm}] \\
      \hline
0& 0.136& 15.24& 222.5& 26.8& 34.1& 32.8& 0.082& 0.118\\
1& 0.136& 15.22& 222.4& 26.7& 35.9& 34.7& 0.087& 0.123\\
2& 0.136& 15.21& 222.2& 26.7& 36.8& 35.6& 0.089& 0.125\\
1r& 0.148& 15.48& 227.7& 27.1& 30.9& 29.6& 0.078& 0.116\\
3& 0.136& 15.21& 216.5& 26.7& 34.7& 33.4& 0.088& 0.124\\
3n& 0.137& 15.20& 218.2& 30.0& 29.3& 16.7& 0.068& 0.107\\
3nr& 0.147& 15.44& 222.9& 31.0& 31.2& 15.5& 0.068& 0.107\\
E& 0.136& 15.28& 223.1& 29.7& 68.2& 66.9& 0.159& 0.174\\
Er& 0.147& 15.53& 228.1& 30.6& 70.2& 68.9& 0.161& 0.176\\[1.5ex]
En& 0.136& 15.27& 222.9& 30.1& 29.1& 66.1& 0.152& 0.172\\
Enr& 0.147& 15.53& 228.2& 31.1& 31.1& 68.3& 0.156& 0.174
    \end{tabular}
  \end{ruledtabular}
  \caption{\label{table:NEDF_B} Saturation, symmetry, and neutron skin
    properties for the various NEDFs.  All values in \si{MeV} unless
    otherwise specified.}
\end{table}

\subsection{Saturation and Symmetry Properties}%
\label{sec:satur-symm-prop}
The isoscalar parameters $j=0$ and quadratic isovector parameters
$j=1$ ($\beta^2$) may be directly related to the saturation and symmetry
properties respectively by expanding the energy per nucleon of
homogeneous nuclear matter Eq.~\eqref{eq:inm} about the symmetric
saturation point $\rho_n = \rho_p = \rho_0/2$:
\begin{gather}
  \frac{\mathcal{E}(\rho_n, \rho_p)}{\rho}
  = \varepsilon_0(\rho) + \varepsilon_2(\rho)\beta^2 +
  \varepsilon_4(\rho) \beta^4 + \order(\beta^6).
\end{gather}
The saturation density $\rho_0$, energy per nucleon $\varepsilon_0$,
and incompressibility $K_0$ are then defined by the minimum
$\varepsilon_0'(\rho_0) = 0$, and depend only on the $j=0$ isoscalar
parameters $a_0$, $b_0$, and $c_0$.  Expanding about $\rho_0$ in
$\delta = (\rho - \rho_0)/3\rho_0$ and in powers of $\beta = (\rho_n -
\rho_p)/\rho$, one can define various ``local'' contributions to the
symmetry energy $S_{2,4}$, its density dependent slope $L_{2,4}$,
etc.:
\begin{gather}
  \label{eq:satsym}
  \setlength{\arraycolsep}{0pt}
  \renewcommand{\arraystretch}{1.2}
  \begin{array}{r >{{}}c<{{}} l >{{}}c<{{}} l >{{}}c<{{}} l >{{}}c<{{}} l}
    \varepsilon_0(\rho) &=&
    \varepsilon_0 && &+& \frac{1}{2}K_{0} \delta^2 &+& \order(\delta^3),\\
    \varepsilon_2(\rho) &=&
    S_{2} &-& L_{2}\delta &+& \frac{1}{2}K_{2} \delta^2 &+& \order(\delta^3),\\
    \varepsilon_4(\rho) &=&
    S_{4} &-& L_{4}\delta &+& \frac{1}{2}K_{4} \delta^2 &+& \order(\delta^3)
  \end{array}
\end{gather}
Since we include also quartic terms $\beta^4$, we must differentiate
between these local symmetry parameters $S_{2}$, $L_{2}$, etc\@. and
the full symmetry parameters defined as the difference between
symmetric matter and pure neutron matter (see also the discussion of
\textcite{Lattimer:2014a}):
\begin{align}
  \label{eq:sym}
  S &= \frac{\mathcal{E}(\rho_0,0)-\mathcal{E}(\rho_0/2,\rho_0/2)}{\rho_0}, \\
  L &= \left.3\rho \diff{}{\rho} \left(
      \frac{\mathcal{E}(\rho,0)}{\rho}
    \right) \right |_{\rho_0} = 3\rho_0 \epsilon_n'(\rho_0). \label{eq:LL}
\end{align}
where $\epsilon_n{\rho}$ is the neutron
equation of state~\eqref{eq:NM}.
Since the saturation density $\rho_0$ minimizes the energy of
symmetric matter, the slope of the full symmetry energy $L$ at
$\rho_0$ depends only on the equation of state of pure neutron matter.
Thus, the \gls{QMC} neutron equation of state alone fixes the global
density dependence of the symmetry energy
$L = 3\rho_0 \epsilon_n'(\rho_0) \approx \SI{30}{MeV}$.   (Small
variations in $L$ seen in Table~\ref{table:NEDF_B} for functionals
fitting neutron matter are due to the variations in $\rho_0$ from fit
to fit.)
\begin{figure}[tb]
  \includegraphics[width=\columnwidth]{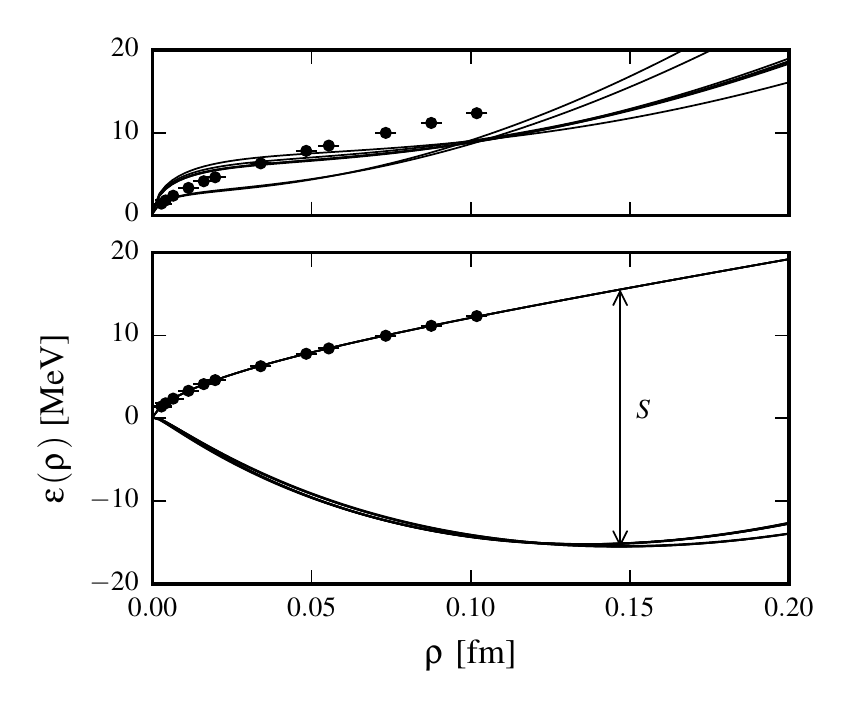}
  \caption{\label{fig:INM}%
    (Color online) Equation of state of symmetric
    nuclear matter (all functionals) and neutron matter
    (for NEDF-3n, 3nr, En, Enr only) in the lower panel
    and pure neutron matter (upper panel)  for NEDF-0, 1, 1r, 2, 3, E, and Er
    (which do not constrain the neutron equation of state and have only $\beta^2$ contributions).
    The black dots shows the \gls{QMC} results~\cite{Wlazowski:2014a}. The
    symmetry energy $S$ is indicated for the
    functionals NEDF-Er and Enr.  The slope $L\approx
    \SI{30}{MeV}$ is fixed by the neutron matter equation
    of state alone (if used as a constraint, see Eq.~\eqref{eq:LL}). In this case the slope
    $L_2/3\rho_0$ may be tuned
    without significantly affecting the mass fit by adjusting the
    insensitive combination $a_1 - b_1\rho_*^{1/3}$, see Section \ref{sec:PCA}. Functionals with
    only quadratic isospin contributions ($\beta^2$) appear to cross near $\rho_n
    \approx \SI{0.1}{fm^{-3}}$ (upper panel).}
\end{figure}

When only $\beta^2$ isospin contributions are included in the
functional, our fits to the nuclear binding energies display a feature
reported in other \glspl{NEDF}: the energy per neutron in pure neutron
matter appears to be well constrained at a density of $\rho_n\approx
\SI{0.1}{fm}^{-3}$ where all functionals cross, see
Fig.~\ref{fig:INM}.  However, the value for the energy per neutron
$\approx \SI{9}{MeV}$ at this point in our fits is significantly
smaller than the value $\approx \SI{12.35}{MeV}$ obtained in QMC
calculations of \textcite{Wlazowski:2014a}.  This feature is not
present when the $\beta^4$ terms are included (NEDF-3n, NEDF-3nr,
NEDF-En and NEDF-Enr)
and the QMC results are reproduced without significantly affecting the
mass fits.  The statement often made in the literature (see
\textcite{Horowitz:2014a} and references therein) that the value of
the symmetry energy at $\rho \approx \SI{0.1}{fm}^{-3}$ is well
constrained by nuclear masses must only be applied to the local
expansion $S_2$ at this density, but not to the symmetry energy
difference $S$ between symmetric and pure neutron matter.

\begin{figure*}[htb]
  \includegraphics[width=\textwidth]{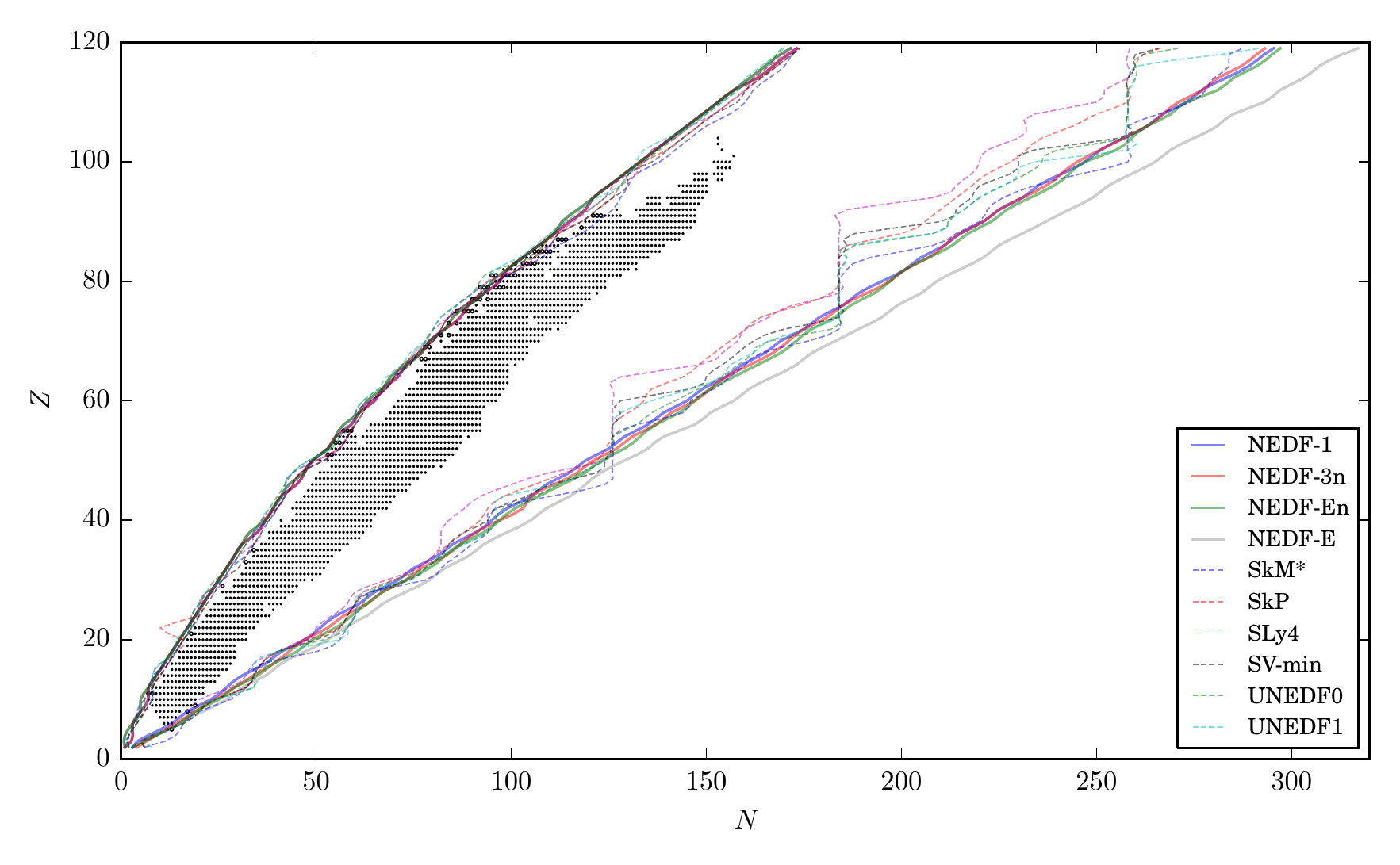}
  \caption{\label{fig:driplines}%
    (Color online) The proton and neutron driplines evaluated with
    \gls{NEDF}-1, \gls{NEDF}-3n,  \gls{NEDF}-E and NEDF-En
    compared with those
    predicted by fully self-consistent calculations using SkM*, SkP,
    SLy4, SV-min, UNEDF0, and UNEDF1 ~\cite{Erler:2012}.  The 2375
    nuclear masses from~\cite{Audi:2012, Wang:2012} are displayed as
    dots and nuclei unstable with respect to the emission of a single
    proton or neutron are denoted with open circles. NEDF-E leads to
    a neutron dripline shifted quite a bit away from the other predictions.
    This is due to changing $L_2$ to a value of the  neutron skin thickness
    in $^{208}$Pb consistent with experiment, see Table~\ref{table:NEDF_B},
    but inconsistent with the neutron equation of state and the values for
    $S$ and $L$, which appear to control the position of the dripline.}
\end{figure*}

\subsection{Symmetry Energy and Neutron Skin Thickness}
As shown in Table~\ref{table:NEDF_B}, the binding energy of nuclear
matter and the symmetry energy predicted by all \glspl{NEDF} fits
agrees well with the value obtained with the mass
formula~\eqref{eq:masses}.  Our fits generally estimate the slope of
the symmetry energy $L_2$ from \SIrange{29}{36}{MeV}, but NEDF-E,
NEDF-En, NEDF-Er, and  NEDF-Enr demonstrate that this is essentially unconstrained by the
masses and can be adjusted independently with the combination $a_1 -
b_1\rho_*^{1/3}$.

We also compute the neutron skin thickness of \ce{^{48}Ca} and
\ce{^{208}Pb}, for which precision measurements
CREX and PREX\footnote{Proposals and related information available at
  \url{http://hallaweb.jlab.org/parity/prex}} are underway,
see~\cite{Horowitz:2014} for details.  For NEDF-1 through NEDF-3rn,
the neutron skin of \ce{^{208}Pb} ranges from \SIrange{0.07}{0.09}{fm}
while the \ce{^{48}Ca} skin ranges from \SIrange{0.106}{0.125}{fm}.
The \ce{^{208}Pb} neutron skin appears quite a bit thinner than a recent
measurement of \SI{0.15(3)}{fm}~\cite{Tarbert:2014x}, but the results
of NEDF-E, NEDF-Er, NEDF-En, and NEDF-Enr demonstrate that this is also controlled by the
same combination $a_1 - b_1\rho_*^{1/3}$ as $L_2$, and hence
unconstrained by the masses.

Since the slope of the symmetry energy $L = L_2 + L_4 + \cdots \approx
\SI{30}{MeV}$ is fixed by the neutron matter equation of
state, requiring a larger value of $L_2 \approx \SI{60}{MeV}$ to
explain the neutron skin thickness of \ce{^{208}Pb} also suggests that
at least quartic terms are required in the functional.

\subsection{Neutron Drip Line}
It is interesting to compare the limits of nucleon stability predicted
by our \gls{NEDF}s. In Fig.~\ref{fig:driplines} we compare the proton
and neutron drip lines obtained with our \gls{NEDF}-1, 3n, E, and En
agains the predictions of UNEDF0 and UNEDF1, as well as those obtained
with other Skyrme parametrizations extracted from the supplemental
data of \textcite{Erler:2012}.  It is a bit unexpected that NEDF-1
and NEDF-3n and En have almost identical neutron driplines considering that
NEDF-1 and E have only quadratic isovector contributions, while NEDF-3n and En also
have quartic terms constrained by the neutron matter equation of state.
Apparently these quartic terms -- and therefore the neutron equation
of state -- play a minor role even in nuclei with a very large neutron
excess $N-Z \approx 160$ when $(N-Z)/A \approx 0.4$ since $[(N-Z)/A]^4
\approx 0.0256$ is small.

The proton driplines are in good agreement with other \glspl{NEDF} (up
to odd-even effects), even though our NEDFs lacks quantum effects.
Our \glspl{NEDF} suggest that the neutron dripline may
extend somewhat further than for conventional functionals.
If significantly more nuclei are stable against nucleon decay, this
may dramatically impact the astrophysical $r$-process, which is
predicted to follow lines of constant separation energy in close
proximity to the neutron dripline~\cite{Meyer:1989, Langake:2001}.
 \textcite{Meyer:1989} considered neutron star ejecta
as the site of r-process nucleosynthesis, and determined that the
reaction flow is very close to the dripline. Even though his
simulations were performed for relatively cold matter (recent
simulations seem to indicate that the star material is somewhat
heated~\cite{Goriely:2011, Rosswog:2014}), it will be interesting to
simulate the r-process using the present \glspl{NEDF}.
There are almost \num{9000} nuclei between the proton and neutron
driplines in case of NEDF-1 and NEDF-3n,
most of which might be stable against nucleon
decay (quantum effects are neglected in this estimate).  This number
is noticeably larger than the estimate
\num[separate-uncertainty]{6900\pm 500} obtained in
Ref.~\cite{Erler:2012}.  The functionals NEDF-E and NEDF-Er lead to a
neutron dripline which is shifted much further, but when the neutron
equation of state is incorporated in NEDF-En and NEDF-Enr, the
neutron dripline is again very close to the previous position given by
NEDF-1 and NEDF-3n. Unlike the neutron skin thickness, which is controlled
by $L_2$, the position of the neutron dripline appears to be controlled by the full
symmetry energy  $S$ and its density dependence $L$.

The predicted position of the neutron dripline will likely affect
the structure of the neutron star crust inferred from older
studies~\cite{Baym:1971yq, Negele:1973, Bulgac:2001y, Bulgac:2002x,
  Magierski:2004x, Magierski:2004a, Magierski:2003, Magierski:2002a}.
The corresponding increase in the neutron skin thickness will also
affect the profile and the pinning energy of quantized vortices in the
neutron star crust~\cite{Avogadro:2007, Avogadro:2008,
  Pizzochero:1997, Pizzochero:2007, Pizzochero:2011, Bulgac:2013a,
  Yu:2003a}.

Fusion cross sections~\cite{Gasques:2005,Adelberger:2011} will also be
significantly altered, particularly in stellar environments where
neutron rich nuclei fuse via pycnonuclear reactions~\cite{Schram:1990,
  Afanasjev:2012}, and where the neutron gas surrounding nuclei leads
to their swelling~\cite{Umar:2015}. A thicker neutron skin with
further enhance this effect.

\subsection{Charge Radii}
Using the parameters determined from the mass fits, the \gls{NEDF} also models
the neutron and proton densities in the nuclei, allowing us to extract the
charge radii for the 883 nuclei in~\cite{Angeli:2013} that intersect with the
2375 nuclear masses we fit from~\cite{Audi:2012, Wang:2012}.  Without any
further adjustment, the \gls{NEDF} gives an \gls{rms} residual for these charge
radii of $\chi_r=\SI{0.135}{fm}$.  Including them in the fit (NEDF-1r, NEDF-3nr, and NEDF-Enr)
improves this to $\chi_r=\SI{0.05}{fm}$ without significantly affecting the
quality of the mass fits.  To compare, fully self-consistent \gls{RMFT} models
realize $\chi_r$ from \SIrange{0.023}{0.041}{fm}~\cite{Agbemava:2014,
  Niksic:2011} depending on the set of nuclei examined.

\begin{figure}[tbp]
  \includegraphics[width=0.9\columnwidth]{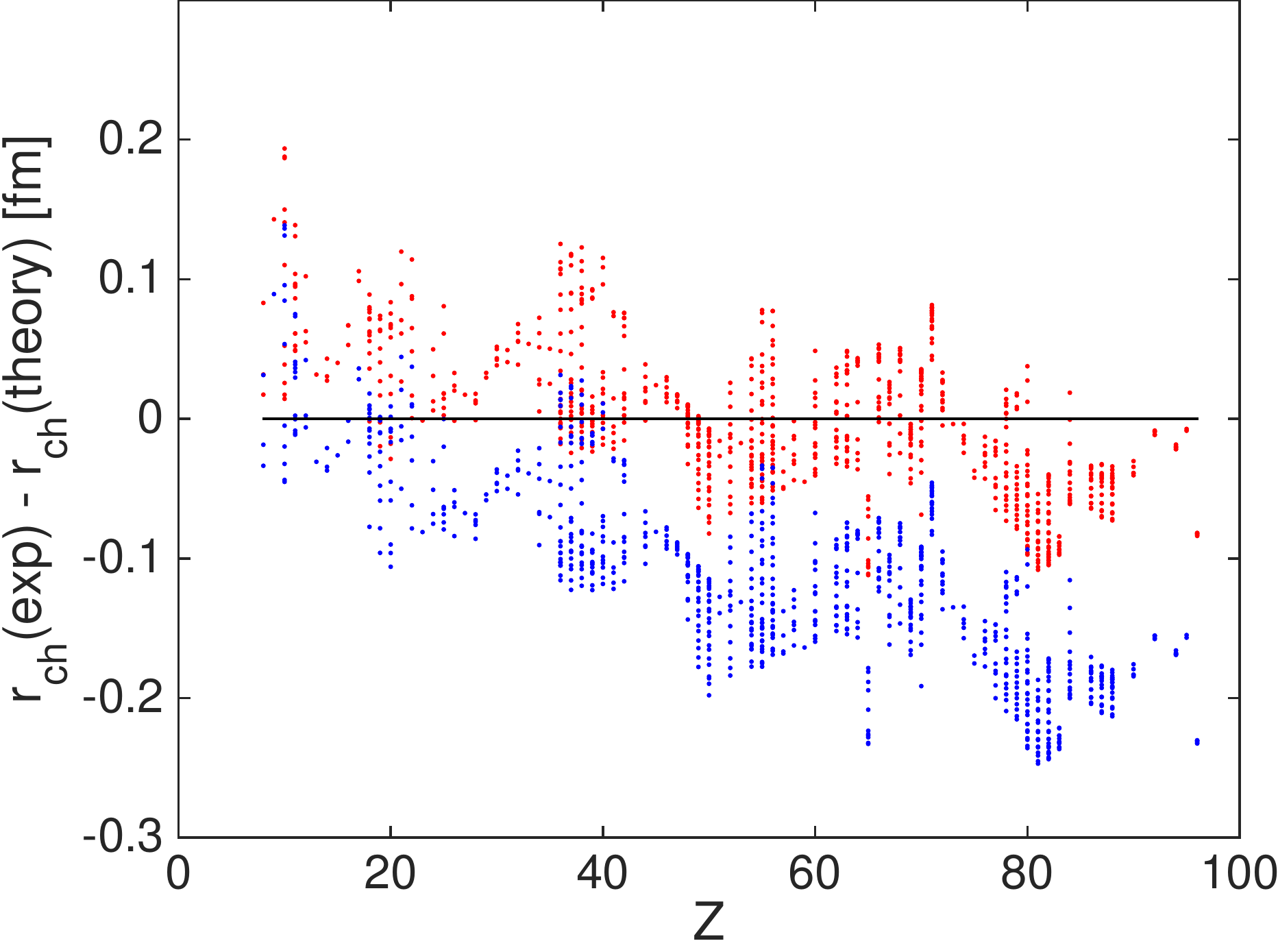}
  \caption{\label{fig:chrad}%
    (Color online) The difference between the
    measured~\cite{Angeli:2013} and computed charge radii of 883
    nuclei evaluated with NEDF-1 (blue) and NEDF-1r (red), see
    Table~\ref{table:liquid_drop}.
    The energy rms corresponding to the blue and red
    points are \SI{2.58}{MeV} with NEDF-1 and \SI{2.71}{MeV} with
    NEDF-1r.  The charge radii rms are in these cases \SI{0.135}{fm}
    with NEDF-1 and \SI{0.051}{fm} with NEDF-1r respectively.  }
\end{figure}

\begin{figure}[tbp]
  \includegraphics[width=0.9\columnwidth]{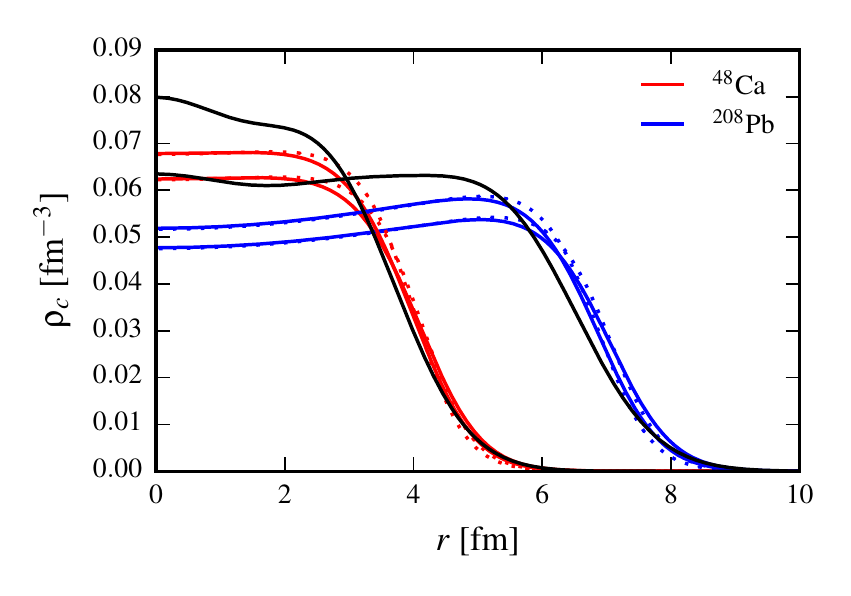}
  \caption{\label{fig:chrad_exp}%
    (Color online) The calculated proton $\rho_p(r)$ (dotted)
    and charge $\rho_{\text{ch}}(r)$ (solid)
    densities for $^{48}$Ca (red) and $^{208}$Pb (blue), calculated
    with NEDF-En (lower curves) and NEDF-Enr (upper curves) compared
    to charge densities (black) extracted from electron scattering
    experiments~\cite{Vries:1987}. }
\end{figure}

\begin{figure}[tbp]
  \includegraphics[width=0.9\columnwidth]{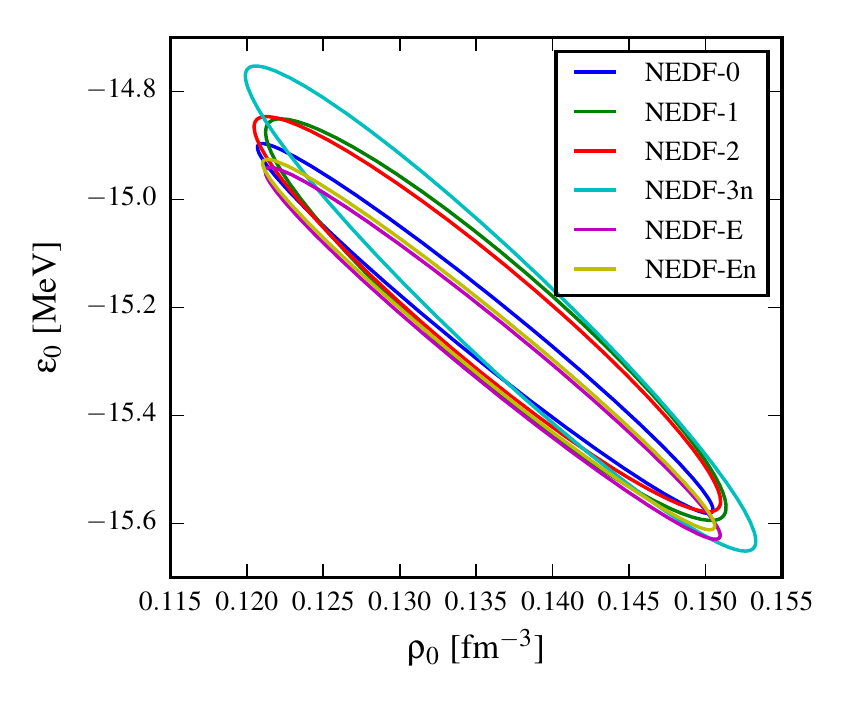}
  \caption{\label{fig:ellipses}%
    (Color online) The various ellipses show the region in the $(\varepsilon_0,\rho_0)$
   plane, in which the NEDF parameters can be changed and to lead to
   changes in the residual
   $\delta \chi_E< \SI{0.2}{MeV}$. While the equilibrium energy $\varepsilon_0$
   and density $\rho_0$ are controlled mainly by the combination
   $\tilde{b}_0+\tilde{c}_0$, which is constrained with very high precision,
   the combination $\tilde{b}_0-\tilde{c}_0$ remains significantly less constraint,
   see Section~\ref{sec:PCA}. This aspect allows us to manipulate to a certain degree
   the saturation properties, while affecting the overall fit only slightly.  }
\end{figure}

The differences between the calculated and measured charge
radii~\cite{Angeli:2013} are illustrated in Fig.~\ref{fig:chrad}.  In
Fig.~\ref{fig:chrad_exp}
we compare the proton and charge densities of $^{48}$Ca and $^{208}$Pb
calculated with NEDF-E and NEDF-Er with the the charge densities
extracted from electron scattering experiments~\cite{Vries:1987}. The
calculated charge density $^{208}$Pb has a slightly larger radius and
slightly smaller diffuseness compared to those extracted from data.
There may be several ways to cure this deficiency. The nuclear
saturation density is not well constrained by the mass fits, and our
values appear somewhat low.  However, one can constrain the
equilibrium density to a value closer to
$\rho_0=\SI{0.16}{fm^{-3}}$ without affecting the mass fit greatly,
see Fig.~\ref{fig:ellipses}. The surface properties are
controlled by the parameter $\eta$, and although a single parameter is
sufficient to fit the masses, one might introduce a density dependence
$\eta \rightarrow \eta_0+\eta_1\rho^{1/3}+\ldots$ or even an isospin
density dependence to better fit the densities.

We do not have yet a clear understanding of why the light nuclei have
slightly smaller radii, while the large nuclei show the opposite
behavior, and whether this aspect is correlated with the nuclear
surface diffuseness.  Within the present approach all nuclei are
spherical; quantum effects may have a noticeable impact on the
calculated charge radii, but so far we do not know how to disentangle
the roles of the bulk and shell-correction effects on these values as
cleanly as we do in the case of binding energies.

In order to compare our results with the experimental values of the
charge radii, we must correct for the motion of the center of mass,
but we find these corrections to be small. In mean-field--like
calculations, the center-of-mass of the system moves in the
self-consistent potential and one needs to account for these
fluctuations before comparing with experimental data. For small
nuclei, it is common to treat these center-of-mass fluctuations by
assuming that the center-of-mass sits in a harmonic oscillator
potential, with the frequency related to the average curvature of
self-consistent nuclear mean-field potential~\cite{Lipkin:1958,
  Negele:1970}.  The calculated charge radius would then be smeared by
convolving with the gaussian center-of-mass wavefunction, and the
correction implemented as a de-convolution which amounts to the
subtraction
\begin{gather}
  \langle r^2_{c}\rangle = \int \d^3{r}\; r^2\rho_{c}(\vect{r})
  - \frac{3}{2}\frac{\hbar}{Am\omega} \\
  \approx \int \d^3{r}\;
  r^2\rho_{c}(\vect{r}) - A^{-2/3}\SI{1.8}{fm^2},\label{eq:lipkin}
\end{gather}
where $\hbar \omega\approx \SI{1.85}{MeV} + A^{-1/3} \SI{35.5}{MeV}$
approximates the harmonic-oscillator energy~\cite{Negele:1970}.  (In
the full calculation, the charge form factors are included in a
similar manner by subtracting the proton and neutron charge radius.)
A similar correction of $\tfrac{3}{4}\hbar\omega$ should be subtracted
from the binding energy of the nucleus.  This prescription is expected
to be accurate enough for light nuclei, but it is not clear that the harmonic
approximation of the mean-field potential will be valid for large
nuclei, which saturate to approximately $\rho_0$ in the core. (Due to the effect
of the nuclear surface tension the nuclei are compressed a bit, but
at the same time Coulomb repulsion tends to deplete the
central charge density.) Here we argue that,
for large nuclei, this correction due to the center-of-mass fluctuation
is small enough to be discarded (an
argument made also by \textcite{Negele:1970} for large nuclei).

To simplify the following argument, we neglect the proton-neutron mass
difference; in actual
calculations we use the experimentally measured nucleon masses.  The intrinsic
radius of a nucleus $r_{\text{rms}}$ is the expectation value of the
square of the distance of the nucleons from the center of mass:
\begin{align}
  r_{\text{rms}}^2 = \frac{1}{A}\sum_{k=1}^{A}\braket{(\vect{r}_k - \vect{R})^2}
  & = \frac{1}{A}\sum_{k=1}^{A}\braket{\vect{r}_k^2} - \braket{\vect{R}^2}
  \nonumber\\
  & = r^2_{m} - \braket{\vect{R}^2}
\end{align}
where $\vect{R} = \sum_{k}\vect{r}_k/A$ is the center-of-mass coordinate and
$\sqrt{r^2_{m}}$ is the \gls{rms} mass radius of the nucleus calculated
directly within the meanfield theory.  This makes it clear that the
center-of-mass correction -- the second term -- is a measure of the size of the
center-of-mass fluctuations
\begin{gather}
  \braket{\vect{R}^2} =
  \frac{1}{A^2}\left \langle \sum_{k=1}^A\vect{r}^2_k
    + \sum_{k\ne l}^A \vect{r}_k\cdot \vect{r}_l \right \rangle  \nonumber \\
  =\frac{r^2_{m}}{A} +
  \frac{1}{A^2}\Braket{\sum_{k\ne l}^A \vect{r}_k\cdot \vect{r}_l}.
  \label{eq:cm0}
\end{gather}
The last term in Eq.~\eqref{eq:cm0} is the value of a two-body observable and
in Hartree-Fock approximation only the exchange term contributes. Using the
known expression for the two-body density in the Hartree-Fock approximation
\begin{gather}
  \rho_2(x,y) = \rho_1(x,x)\rho_1(y,y) - \rho_1(x,y)\rho_1(y,x),
\end{gather}
where $\rho_1(x,y) =
\rho_1(\vect{r}_1,\sigma_1,\tau_1,\vect{r}_2,\sigma_2,\tau_2)$ is the one-body
density matrix.  We now approximate this using the Thomas-Fermi approximation
for the density matrix
\begin{gather}
  \rho_1(x,y)=\delta_{\sigma_1,\sigma_2}\delta_{\tau_1,\tau_2}
  \int_{k\le k_F} \frac{\d^3{k}}{(2\pi)^3}
  \exp\left( \I\vect{k}\cdot \vect{s}\right), \label{eq:TF}
\end{gather}
where $k_F$ is the local neutron or proton Fermi momentum and $\vect{s} =
\vect{r}_1-\vect{r}_2$, one can show that
\begin{gather}
  \frac{1}{A^2}\Braket{\sum_{k\ne l}^A \vect{r}_k\cdot \vect{r}_l}
  = -\frac{r^2_{m}}{A} + \order(A^{-1}).\label{eq:cm1}
\end{gather}
Thus, the leading order terms of the direct and exchange contributions
cancel exactly leaving $\braket{\vect{R}^2} = \order(A^{-1})$, which
is significantly smaller than the correction suggested by the harmonic
oscillator assumption~\cite{Lipkin:1958, Negele:1970}.  We therefore
neglect these correction in our fits as they are insignificant in
comparison with missing shell effects etc.

It is instructive to perform a similar calculation for the kinetic energy of
the center-of-mass
\begin{align}
  E_{\text{cm}} &= \frac{1}{2mA} \left \langle \vect{P}^2 \right \rangle
  = \frac{1}{2mA} \Braket{\left( \sum_{k=1}^A \vect{p}_k \right)^2}\\
  &= \frac{\hbar^2}{2mA} \Braket{
    \sum_{k=1}^A \abs{ \grad \psi_k(\vect{r})}^2
  }
  +
  \frac{1}{2mA} \Braket{
    \sum_{k \ne l} \vect{p}_k \cdot \vect{p}_l
  }, \nonumber
\end{align}
(For clarity we have suppressed spin and isospin indices.)  Using again the
Thomas-Fermi approximation~\eqref{eq:TF}, and including the Weizsäcker
correction~\cite{Weizsacker:1935, Dreizler:1990lr, Brack:1997} to the kinetic
energy density in the direct term, we determine that
\begin{gather}
  E_{\text{cm}} \sim \frac{\hbar^2}{2mA}
  \int \d^3{r} \left(
    \grad\rho^{1/2}(\vect{r})
  \right)^2
  \sim \frac{\hbar^2}{2mr_0^2} A^{-1/3}
\end{gather}
where $m$ is a nucleon mass, $R = r_0A^{1/3}$ is the nuclear radius, and
$\hbar^2/2mr_0^2 \sim \SI{35}{MeV}$ is of the order of the Fermi energy.

The arguments presented here on the role of center-of-mass fluctuations
are based on a Hartree-Fock-like approach, since we need to evaluate
two-body observables, and strictly speaking within DFT, one does not
have access to the two-body densities.

\begin{figure}[htp]
  \includegraphics[width=\columnwidth]{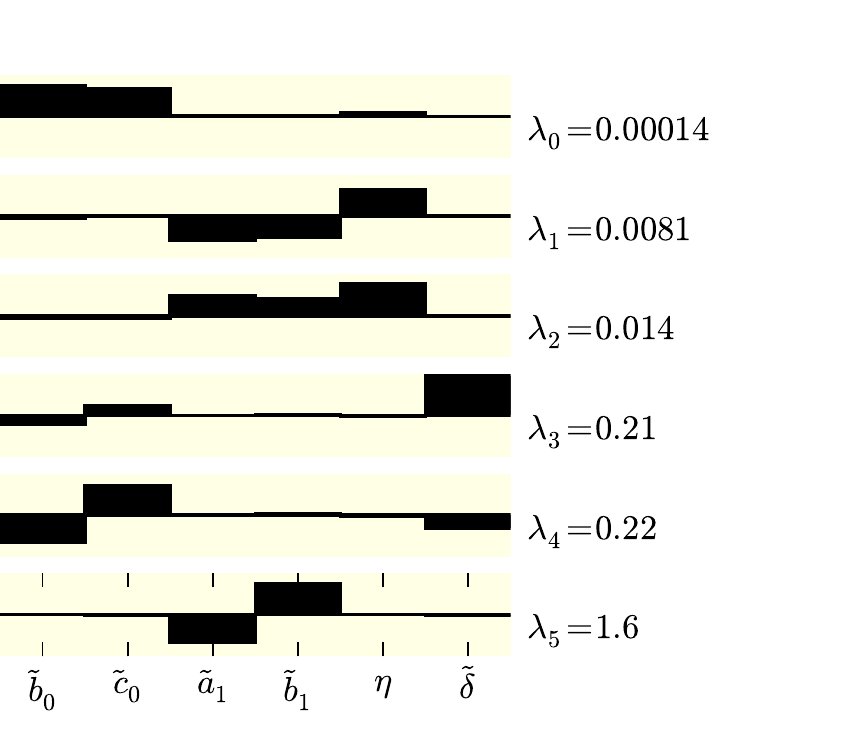}
  \includegraphics[width=\columnwidth]{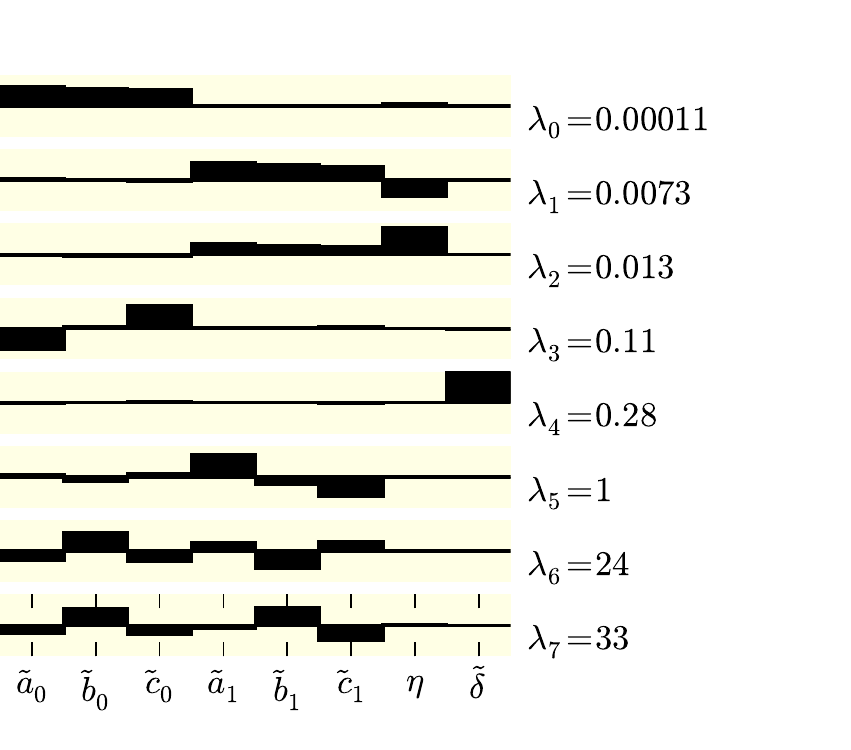}
  \caption{\label{fig:PCA}%
    (Color online) Principal component analysis for the NEDF-1 fit
    (top) and the NEDF-3 fit (bottom).  Plotted are the components of
    the eigenvectors $\vect{v}_n$ defining the principal component
    Eq.~\eqref{eq:pca_p} as linear combinations of the various
    dimensionless parameters. From this we see that the
    most-significant component $p_0 \approx \tilde{b}_0 + \tilde{c}_0$
    which fixes the saturation energy to high precision.
    At the same time the component $p_4\approx \tilde({b}_0-\tilde{c}_0$
    in NEDF-1 (and similarly in NEDF-3n) is not well constrained.   We also see
    that the least-significant component $p_{5} \approx \tilde{a}_1 -
    \tilde{b}_1$ is essentially unconstrained.  For NEDF-3, we find
    three insensitive components, two of which can be used to set the
    smallest parameters $a_0=c_1=0$.  After removing these, one
    obtains a similar analysis as for NEDF-1 above.  }
\end{figure}

\subsection{Principal Component Analysis}
\label{sec:PCA}

The parameters listed in Table~\ref{table:NEDF_A} are highly
correlated.  To analyze these, we consider as significant changes
$\delta\chi_E \approx \SI{0.1}{MeV}$ since this is the typical level
of sensitivity of the mass fits.  We keep the changes relatively small
because otherwise the model is not well approximated by a quadratic
error model if $\delta \chi_E > \SI{0.1}{MeV}$.  Numerically we find
that even \SI{0.1}{MeV} is too large, but yields qualitatively correct
information after a full refitting.  Note that $\delta(\chi_E^2) =
(\chi_E + \delta\chi_E)^2 - \chi_E^2 = 2\,\chi_E\, \delta\chi_E +
(\delta\chi_E)^2$, so we must normalize $\delta(\chi_E^2)$ by
$2\,\chi_E \cdot \SI{0.1}{MeV}$ in order to consider changes
$\delta\chi_E \approx \SI{0.1}{MeV}$.

To compare the parameters in a meaningful way, we must make them
dimensionless and of order unity.  We do this as in
Table~\ref{table:NEDF_A} by scaling with appropriate powers of $\rho_*
= 0.15\si{fm^{-3}}$ and $\epsilon_F = \tfrac{\hbar^2}{2m}
(3\pi^2\rho_*/2)^{2/3} = \SI{35.29420}{MeV}$, which we take as fixed
parameters close to the saturation values:
\begin{align}
  \tilde{a}_j &= \frac{a_j\rho_*^{2/3}}{\epsilon_F}, &
  \tilde{b}_j &= \frac{b_j\rho_*}{\epsilon_F}, &
  \tilde{c}_j &= \frac{a_j\rho_*^{4/3}}{\epsilon_F}.
  \label{eq:dimensionless}
\end{align}
(It is important to retain a significant number of digits for isoscalar
quantities, as it will be come clearer below.)
In particular, we consider the covariance matrix $\mat{C}$ such that
the residual deviation is
\begin{subequations}
  \begin{gather}
    \frac{\delta(\chi_E^2)}{2\chi_E \cdot \SI{0.1}{MeV}}
    \approx
    \vect{\delta}^T\cdot\mat{C}^{-1}\cdot\vect{\delta}
    = \sum_{n}\frac{(\delta p_n)^2}{\lambda_n^2}.
  \end{gather}
  where $\vect{\delta}$ is the deviations vector of the dimensionless
  parameters Eq.~\eqref{eq:dimensionless} from their best fit
  values as listed in Table~\ref{table:NEDF_A}, and we have
  diagonalized $\mat{C}\vect{v}_n = \lambda_n^2\vect{v}_{n}$ to obtain
  the principal components $p_n$
  \begin{gather}
    \label{eq:pca_p}
    p_n = \vect{v}_n
    \cdot\begin{pmatrix}
      \tilde{a}_0&
      \tilde{b}_0&
      \cdots&
      \eta&
      \tilde{\delta}
    \end{pmatrix}.
  \end{gather}
\end{subequations}
Since the parameters are of order unity, we may directly consider the
$\lambda_n$ as a measure of the errors: changing $p_n$ by $\lambda_n$
will affect the fit on the scale of $\delta\chi_E \approx
\SI{0.1}{MeV}$. Therefore, the smaller the value of the parameter
$\lambda_n$, the more precisely the fit to nuclear masses constrains
the value of the corresponding linear combination of \gls{NEDF}
parameters.

The principal components for fits NEDF-1 and NEDF-3 are shown in
Fig.~\ref{fig:PCA}.  Both show similar features that can be
understood in terms of the saturation and symmetry parameters
Eqs~\eqref{eq:satsym} which we
may express as:
\begin{subequations}
  \begin{alignat}{3}
    \frac{\varepsilon_0}{\epsilon_F} &= \tfrac{3}{5} + \tilde{a}_0
    +  &&\tilde{b}_0\; + &&\tilde{c}_0,\label{eq:e0}\\
    0 &= \tfrac{3}{5} + \tilde{a}_0
    + & \tfrac{3}{2}&\tilde{b}_0\; + &2&\tilde{c}_0,\label{eq:rho0}\\
    \frac{K}{-18\epsilon_F} &= \tfrac{3}{5} + \tilde{a}_0 && - &2&\tilde{c}_0,
  \end{alignat}
  \begin{align}
    \frac{S}{\epsilon_F}
    &= \tfrac{3(2^{2/3}-1)}{5}
     + ( \tilde{a}_1 + \tilde{b}_1 + \tilde{c}_1 )
     + ( \tilde{a}_2 + \tilde{b}_2 + \tilde{c}_2 ),
    \label{eq:S}\\
    \frac{L}{\epsilon_F}
    &= \frac{2S}{\epsilon_F} + (\tilde{b}_1 + 2\tilde{c}_1)\;
    + (\tilde{b}_2 + 2\tilde{c}_2).
    \label{eq:L}
  \end{align}
\end{subequations}
The most significant component $p_0$ in both fits is the sum of the
$j=0$ coefficients $\tilde{a}_0 + \tilde{b}_0 + \tilde{c}_0$ which
fixes the saturation energy $\epsilon_0$ Eq.~\eqref{eq:e0} to very
high precision.  Next are mixtures of $\eta$ and the symmetry energy
$S$ Eq.~\eqref{eq:S}, which are correlated by the finite size of the
nuclei; the latter is the sum of the $j=1$ coefficients $\tilde{a}_1 +
\tilde{b}_1 + \tilde{c}_1$ (recall that these fits have no $\beta^4$
components).  Finally, we have some insignificant parameters,
including $\tilde{\delta}$.

In the case of NEDF-3, we see that two parameters are completely
insignificant.  These include $\tilde{a}_0 \approx -0.088$ and $\tilde{c}_1 =
-0.017$.  These values are an order of magnitude smaller than the other
coefficients: hence, the insignificant components can be easily
removed by setting $a_0 = c_1 = 0$ which we do in most of our fits.

Finally, both plots indicate that a combination of the $j=1$
parameters is highly insignificant.  Thus, in NEDF-1, the combination
$\tilde{b}_1 - \tilde{a}_{1}$ can be given almost any value of order
unity without changing $\chi_E$ more than \SI{0.1}{MeV}.  This is
directly tested in NEDF-E, NEDF-Er, NEDF-En, and NEDF-Enr where we
change the sign of $b_1$, setting $a_1 = b_1 \rho_*^{2/3}$.  Indeed,
we see that $\chi_E$ changed by about \SI{0.1}{MeV}.  Notice from
Table~\ref{table:NEDF_B} that the slope of the symmetry energy $L_2$
changes from about $\SI{30}{MeV}$ to $\SI{70}{MeV}$ while the other
parameters remain about the same.  This also significantly changes the
neutron skin thickness, demonstrating a correlation between $L_2$ and
the skin thickness, similar to that seen in other mean-field
models~\cite{Vinas:2014}. This is consistent with Eq.~\eqref{eq:L}
where we see that $\tilde{b}_1$ gives us a direct handle on $L_2$.

\section{Dynamical Properties}%
\label{sec:dynamical-properties}
The NEDF can be used to study a variety of dynamical phenomena,
excitation of nuclei by various external probes and various (large
amplitude) collective modes, low energy collisions of nuclei, induced
fission and fusion, etc.  A dynamical interpretation of the functional
follows by considering a coupled hydrodynamic theory of protons and
neutrons.

Consider first a single species with number density $\rho$
and a local momentum potential $\phi$ such that the local fluid
momentum $\vect{p} = m \vect{v} = \grad\phi$.  The Euler-Lagrange
equations for the following Lagrangian density
\begin{subequations}
  \label{eq:hydro}
  \begin{gather}
    \mathcal{L}
    = -\rho\left(\dot{\phi} + \frac{1}{2m}(\grad\phi)^2\right)
    - \mathcal{E}(\rho)
    - \eta\frac{\hbar^2}{2m}(\grad\sqrt{\rho})^2
  \end{gather}
  define the following hydrodynamic equations:
  \begin{align}
    & \dot{\rho} + \grad(\rho\vect{v}) = 0,\\
    &\dot{\phi} + \frac{mv^2}{2} + \mathcal{E}'(\rho)
    - \eta\frac{\hbar^2}{2m}\frac{\grad^2\sqrt{\rho}}{\sqrt{\rho}} = 0,
  \end{align}
  the latter of which is sometimes expressed as
  \begin{gather}
     m\dot{\vect{v}} + \grad\left[
      \frac{mv^2}{2} + \mathcal{E}'(\rho)
      -\eta\frac{\hbar^2}{2m}\frac{\grad^2\sqrt{\rho}}{\sqrt{\rho}}
    \right] = 0.
  \end{gather}
\end{subequations}
Recognizing that the last term in these equations has the form of a ``quantum
pressure'' term with a modified Planck's constant $\tilde{\hbar} =
\eta^{1/2}\hbar$, one can check that this hydrodynamic theory is equivalent to
the following ``quantum'' theory for a wavefunction
$\psi=\sqrt{\rho}\exp(\I\phi/\tilde{\hbar})=\sqrt{\rho}\exp(\I\theta)$:
\begin{subequations}
  \label{eq:nlseq}
  \begin{align}
    \mathcal{L}(\psi, \dot{\psi}) &= \psi^\dagger\left(
      \I\tilde{\hbar}\partial_t + \frac{\tilde{\hbar}^2\grad^2}{2m}\right)\psi -
    \mathcal{E}(\rho),\\
    \I\tilde{\hbar}\dot{\psi}
    &= -\frac{\tilde{\hbar}^2}{2m}\grad^2\psi + \mathcal{E}'(\rho)\psi.
  \end{align}
\end{subequations}
Note that with these equations, the action is invariant under the Galilean
boost to a frame with velocity $\vect{v}$:
\begin{align}\label{eq:galilean}
  \psi(\vect{x}, t) &\rightarrow e^{\I\theta}
  \psi(\vect{x} - \vect{v}t, t), &
  \tilde{\hbar}\theta &= m\vect{v}\cdot\vect{x} - \frac{mv^2 t}{2}.
\end{align}
Equations~\eqref{eq:nlseq} have exactly the same form as a non-linear
Schrödinger equation that would be used to study the collective
wavefunction of a superfluid \gls{BEC} with homogeneous energy density
$\mathcal{E}(\rho)$.  The implementation of the hydrodynamic equations
as a non-linear Schr\"odinger equation allows for stable numerical
implementations and may be solved efficiently using standard
algorithms. Somewhat surprisingly this ``Schr\"odinger equation''
admits ``quantized vortices'', albeit with a modified ``Planck's
constant'' $\tilde{\hbar}$.  The quantization of vortices here is
purely a formal artifact since it depends on the free parameter
$\eta$.  Although it might be appealing to consider
Eqs.~\eqref{eq:nlseq} as a theory for superfluid dimers comprising two
protons or two neutrons, the best-fit value of $\eta \approx 0.5$ does
not presently admit this interpretation.  As mentioned earlier, to
restore properly quantized vortices requires $\eta = 1/4$ and a
rescaling of the wavefunction, by normalizing it to the total particle
number -- i.e\@. $2\rho = \abs{\psi}^2$ -- so that the factors of
$\eta = 1/4$ cancel in Eqs.~\eqref{eq:nlseq}, yielding a properly
normalized Schrödinger equation.

\begin{subequations}
  One can similarly recast our complete \gls{NEDF} in terms of two ``complex
  wavefunctions'' with a modified ``Planck's constant'' $\tilde{\hbar} =
  \sqrt{\eta}\hbar$:
  \begin{gather}
    \label{eq:psis}
    \psi_{n,p} = \sqrt{\rho_{n,p}}\exp(\I\theta_{n,p}), \quad
    m_{n,p}\vect{v}_{n,p} = \tilde{\hbar} \grad\theta_{n,p},
  \end{gather}
  through the Lagrangian density
  \begin{gather}
    \mathcal{L} = \I\tilde{\hbar}(\psi_n^\dagger\dot{\psi}_n
    + \psi_p^\dagger\dot{\psi}_p) - \mathcal{E}[\psi_n, \psi_p]
  \end{gather}
  where
  \begin{multline}
    \mathcal{E}[\psi_n, \psi_p] =
    \frac{\tilde{\hbar}^2}{2m_n}\grad\psi_n^\dagger\cdot\grad\psi_n
    +
    \frac{\tilde{\hbar}^2}{2m_p}\grad\psi_p^\dagger\cdot\grad\psi_p
    +\\
    + \mathcal{E}_{\text{entrain}}(\psi_n, \psi_p)
    + \mathcal{E}_h(\rho_n, \rho_p),
  \end{multline}
  and the homogeneous energy-density (no gradients) is
  \begin{multline}
    \mathcal{E}_h(\rho_n, \rho_p) =
    \frac{3\hbar^2(3\pi^2)^{2/3}}{10}\left(
      \frac{\rho_n^{5/3}}{m_n}
      +
      \frac{\rho_p^{5/3}}{m_p}
    \right) + \\
    + \tfrac{1}{2}V_C\rho_c - \frac{e^2\pi}{4} \left(
      \frac{3\rho_p}{\pi}\right)^{4/3} \\
    + \sum_{j=0}^{2}(a_j \rho^{5/3} + a_j \rho^{6/3} + a_j \rho^{7/3})\left(
      \frac{\rho_{n} - \rho_{p}}{\rho}
    \right)^{2j} +\\
  \end{multline}
\end{subequations}
The representation of the densities in the form~\eqref{eq:psis} is similar in
spirit to the decomposition of the single-particle density matrix championed in
the theory of large amplitude collective motion $\rho \rightarrow e^{-\I\chi}
\rho e^{\I\chi}$~\cite{Baranger:1978, Brink:1976, Ring:2004}. The energy of a
nucleus is thus also uniquely separated into the collective kinetic energy (the
terms explicitly depending on $\psi_{n,p}$) and the internal energy depends
only on the ``shape'' of a nucleus, with no internal quasi-particle
excitations.

Ignoring for a moment the entrainment term
$\mathcal{E}_{\text{entrain}}(\psi_n, \psi_p)$, dynamics can be implemented
using the following coupled non-linear ``Schrödinger'' equations:
\begin{subequations}
  \begin{align}
    \I\tilde{\hbar}\pdiff{\psi_{n,p}}{t}  &=
    -\frac{\tilde{\hbar}^2}{2m_{n,p}}\grad^2\psi_{n,p}
    + U_{n,p}\psi_{n,p}, \label{eq:tdwfs}\\
    U_{n,p} &= \pdiff{\mathcal{E}_{h}(\rho_n, \rho_p)}{\rho_{n,p}}.
  \end{align}
\end{subequations}

\subsection{Entrainment}
As mentioned above, this theory is covariant under Galilean boosts to a frame
with velocity $\vect{v}$ (Eq.~\eqref{eq:galilean}).  Since both protons and
neutrons experience the same boost $\vect{v}_{n,p} \rightarrow \vect{v}_{n,p} +
\vect{v}$ , we can introduce an additional Galilean invariant term in the
two-component system proportional to $\abs{\vect{v}_n - \vect{v}_p}^2$,
obtaining the collective kinetic energy:
\begin{gather}
  \label{eq:Kinetic}
  \frac{m_n\rho_n v_n^2}{2} + \frac{m_p\rho_p v_p^2}{2}
  + \alpha \frac{m_n m_p \rho_n\rho_p\abs{\vect{v}_n - \vect{v}_p}^2}{2m\rho}.
\end{gather}
The parameter $\alpha$ controls the entrainment of the neutron and proton fluids,
and the form has been chosen so that the term vanishes when either density
vanishes.

As with the quadratic terms above which enter the functional as
$\abs{\grad\psi_{n,p}}^2$, the entrainment term must be constructed
from the wavefunctions rather than from the velocities to avoid singularities
when the density vanishes (i.e\@. in the cores of quantized vortices or in the
tails of the nucleus).  This can be done by noting that the following
``currents'' transform under Galilean boosts as
\begin{align}
  \vect{J}_{n,p} &= -\I\hbar\psi_{n,p}^\dagger\grad\psi_{n,p}, &
  \vect{J}_{n,p} &\rightarrow \vect{J}_{n,p} + m_{n,p}\rho_{n,p}\vect{v}.
\end{align}
Hence, we define the Galilean invariant entrainment term:
\begin{multline}
  \mathcal{E}_{\text{entrain}}(\psi_n, \psi_p) =\\
  = \alpha \frac{m_nm_p\rho_n\rho_p}{2m\rho_0}\Abs{
    \frac{\vect{J}_{n}}{m_n\rho_n} - \frac{\vect{J}_{p}}{m_p\rho_p}}^2\\
  = \alpha \frac{\hbar^2}{2m\rho}
  \Abs{\frac{m_p^{1/2}}{m_n^{1/2}} \psi_p \grad{ \psi_n}
    - \frac{m_n^{1/2}}{m_p^{1/2}} \psi_n \grad \psi_p}^2.
  \label{eq:entrainment}
\end{multline}

For static properties, this term has the form
\begin{multline}
  \mathcal{E}_{\text{entrain}}(\rho_n, \rho_p) =\\
  = \alpha \frac{\hbar^2}{2m\rho}
  \Abs{\frac{m_p^{1/2}}{m_n^{1/2}} \rho_p^{1/2} \grad{\rho_n^{1/2}}
    - \frac{m_n^{1/2}}{m_p^{1/2}} \rho_n^{1/2} \grad \rho_p^{1/2}}^2,
\end{multline}
and leads to a coupling between the neutron and proton density gradients
$\grad\rho_n\cdot\grad\rho_p$.  \textcite{Bodmer:1960} introduced this type
of term, and it also appears in various Skyrme parameterizations of the
\gls{NEDF}. By varying $\alpha$ from $-0.5$ to
$0.5$, $\chi_E$ changed by at most \SI{15}{keV}.  The significant effect of
this term is seen however in the dynamics, where the
motion of one fluid will drag along
the other, affecting strongly the excitation energies of isovector modes
such as the \glspl{GDR}.

Entrainment (the Andreev-Bashkin effect) was predicted by
\textcite{Andreev:1975} to occur in superfluid mixtures of $^3$He and
$^4$He, and is rather surprising at first sight, since superfluids are
expected to flow without resistance. In particular, one might have
expected that if somehow one would bring into motion only one
superfluid component, superfluidity will have the consequence that the
other component remains at rest.  The entrainment
term~\eqref{eq:entrainment} is indeed dissipationless, and thus it
does not violate superfluidity, but allows the motion of one
superfluid to influence (entrain) the other.  It is natural to expect
a similar phenomenon to arise in nuclei, where proton and neutron
(super)fluids can coexist.  Entrainment should also plays a role in
neutron stars and has been studied intermitently since
1975~\cite{Volovik:1975a, Vardanyan:1981, Alpar:1984x, Babaev:2004,
  Gusakov:2005, Chamel:2006, Chamel:2013a}.  The formalism describing
these systems is called three fluid hydrodynamics -- two superfluids
and one normal component -- and is generalization of Landau's two
fluid hydrodynamic phenomenological model of superfluids at finite
temperatures below, but the critical temperature. Since in nuclear
systems both neutron and proton subsystems can have a superfluid and a
normal components at finite temperatures, and since the normal
components can move independently in isovector modes, a proper
generalization of the superfluid dynamics to nuclear systems would be
a four fluid hydrodynamics, with two superfluid and two normal
components, thus a somewhat more complex system than the superfluid
mixtures considered so far in literature.

\subsection{Shell Effects}
\label{sec:shell-effects}

Shell-corrections have a typical amplitude of several \si{MeV}s.  Even though
they represent a relatively small correction to the total energy of the system,
the can have a major effect on low energy nuclear dynamics.  Their magnitude is
controlled by the spin-orbit interaction, the pairing effects, and the nuclear
shape.  There are several prescriptions one can use to compute shell
corrections, but the idea behind them is the same. The shell-corrections are a
function $E_{\text{sc}}(\{\epsilon_k\})$ of single-particle energies $\epsilon_k$
\begin{gather}
  \mat{h}_\tau \psi_{\tau,k} = \epsilon_{\tau,k}\psi_{\tau,k}, \label{eq:speqs}
\end{gather}
where $\psi_{\tau,k}$ are the single-particle wave functions for the
single-particle Hamiltonian $\mat{h}_\tau $ for proton and neutrons $\tau =
n,p$.  This Hamiltonian is the functional derivative of the energy density
(here $\alpha = 0$)
\begin{align}
  & \mathcal{F}[\rho_n,\rho_p]=
  \sum_{k,\tau=n,p}  \frac{\hbar^2}{2m_\tau}
  \Abs{\grad \psi_{\tau,k}}^2  \nonumber \\
  &+ \sum_{\tau=n,p}   \left (\eta -\frac{1}{9} \right )
  \frac{\hbar^2}{2m_\tau}  \Abs{\grad \rho_{\tau}^{1/2}}^2 \nonumber \\
   & + \sum_{j=0}^2\mathcal{E}_j(\rho) \beta^{2j} + \mathcal{E}_{C}(\rho_n, \rho_p)   \nonumber \\
  &  + \mathrm{spin\!-\!orbit + pairing \; contributions.}
  \label{eq:Usc}
\end{align}
The main difference between
this energy density and Eq.~\eqref{eq:NEDF} is that now the $\psi_{\tau,k}$ are
the single-particle neutron and proton wave functions.

There is an ambiguity is separating the contribution of the gradient
terms and other forms of this energy density can be considered. For
example, one can introduce an additional parameter $\xi$ (which might
have a density and even an isospin dependence) and define the energy density
$\mathcal{F}[\rho_n,\rho_p]$ as:
\begin{align}
  & \mathcal{F}[\rho_n,\rho_p]=
  \sum_{k,\tau=n,p} (1+\xi)  \frac{\hbar^2}{2m_\tau}
  \Abs{\grad \psi_{\tau,k}}^2  \nonumber \\
  & -\xi \sum_{\tau=n,p} \frac{\hbar^2}{2m_\tau}
   \tfrac{3}{5}(3\pi^2)^{2/3} \rho_\tau^{5/3}  \nonumber \\
   &   + \sum_{\tau=n,p}   \left (\eta -(1+\xi)\frac{1}{9} \right )
  \frac{\hbar^2}{2m_\tau}  \Abs{\grad \rho_{\tau}^{1/2}}^2 \nonumber \\
   & + \sum_{j=0}^2\mathcal{E}_j(\rho) \beta^{2j} + \mathcal{E}_{C}(\rho_n, \rho_p)   \nonumber \\
  &  + \mathrm{spin\!-\!orbit + pairing \; contributions.}
  \label{eq:Usc1}
\end{align}
The introduction of such a parameter $\xi$ requires adding additional
terms to restore the Galilean invariance of this energy density
functional~\cite{Engel:1975} (strictly speaking the Galilean
invariance of the intrinsic energy of a nucleus), and its introduction
will also lead to a modification of the Thomas-Reiche-Kuhn sum rule
for \gls{GDR}. The semiclassical limit of both Eqs. \eqref{eq:Usc} and
\eqref{eq:Usc1} are identical by construction.

As in the case of the unitary Femi gas~\cite{Bulgac:2007a, Bulgac:2011, Bulgac:2013b},
without additional information form either the spectral weight function or from spectroscopy, the
interpretation of the terms $a_j\rho^{5/3}$ as either being
the semiclassical limit of the appropriate kinetic energy densities or of
interaction energy densities is ambiguous. This ambiguity is related to the
discussion above.

The equations of motion
\eqref{eq:tdwfs} have to be augmented now with the contribution to the potential
arising from $E_{\text{sc}}(\{\epsilon_k\})$ as follows:
\begin{multline}
  \I\tilde{\hbar} \pdiff{\psi_{n,p}}{t} =
  -\frac{\tilde{\hbar}^2}{2m_{n,p}}\laplacian\psi_{n,p} + U_{n,p}\psi_{n,p} \\
  +\sum_l\frac{\partial E_{\text{sc}}(\{\epsilon_k\})}{\partial \epsilon_l}
 \frac{\delta \epsilon_l}{\delta  \rho_{n,p}}  \psi_{n,p}.
  \label{eq:tdwfs_full}
\end{multline}

\begin{figure}[htp]
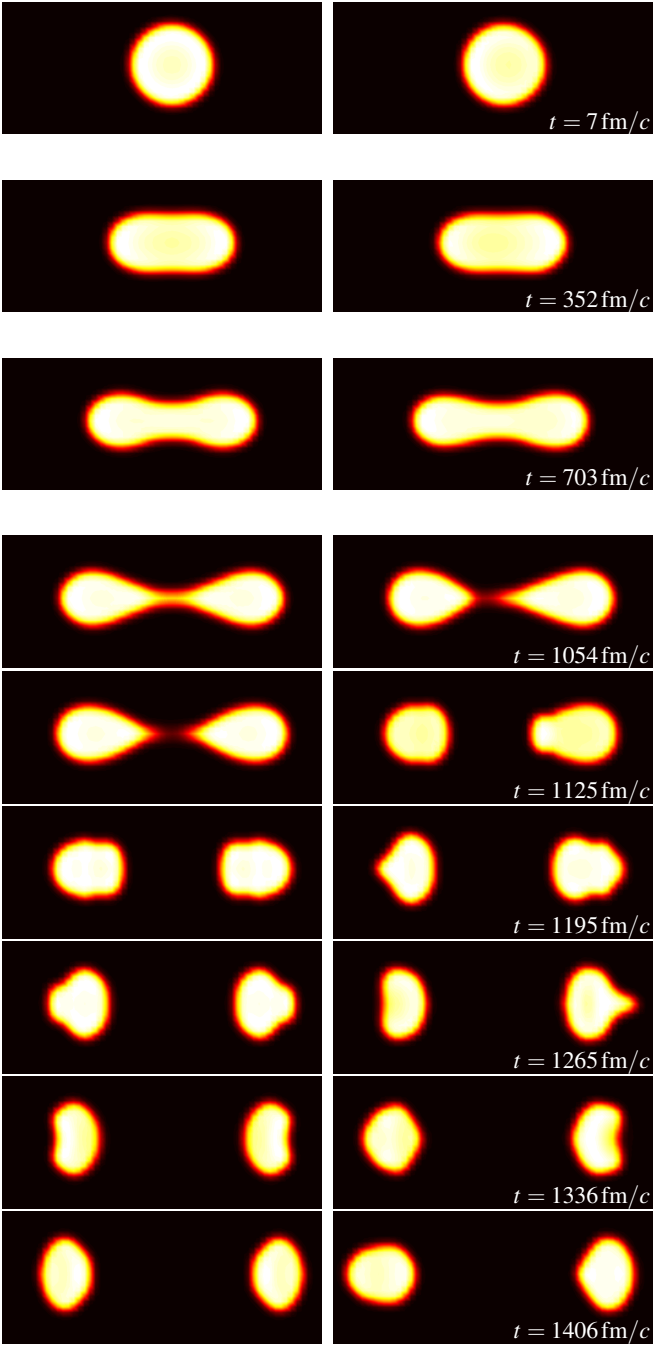

  \def\ig#1{\adjustimage{%
    trim=80 140 80 120, clip=true, width=0.5\columnwidth-0.5em/2}{#1}}
  \let\igg\relax
  \newcommand{\igg}[2][1]{%
    \noindent\ig{sym_#2}\hspace{0.5em}\ig{asym_#2}%
    \adjustbox{raise=2pt}{\llap{\color{white}$t=\SI{#1}{fm}/c\,$}}\\%
  }
  \vbox{
    \baselineskip=0pt
    \parskip=0pt
    \igg[7]{fission1}\vspace{1.8em}
    \igg[352]{fission2}\vspace{1.8em}
    \igg[703]{fission3}\vspace{1.8em}
    \igg[1054]{fission4}
    \igg[1125]{fission5}
    \igg[1195]{fission6}
    \igg[1265]{fission7}
    \igg[1336]{fission8}
    \igg[1406]{fission9}
  }
  \caption{\label{fig:fission}%
    (Color online) Induced fission of $^{238}$U excited with an initial
    quadrupole and octupole potential velocity
    $\theta_n(\vect{r})=\theta_p(\vect{r}) =q_2(2z^2-x^2-y^2)+q_3(5z^2-3)z$
    starting from the ground state, see Eqs.~\eqref{eq:psis}. The symmetric
    case (left) has $q_3=0$ and an excitation energy of \SI{7.68}{MeV}
    (initial collective kinetic energy). The
    asymmetric case (right) has an excitation energy of \SI{10.33}{MeV}
    (initial collective kinetic energy) and the
    final charges, neutron numbers, and atomic masses of the two fragments are
    $Z_1 = 49.3$, $N_1 = 78.4$, $A_1 = 127.7$, $Z_2 = 42.7$, $N_2 = 67.6$, and
    $A_2 = 110.3$.  Both fragments emerge with a significant amount of
    collective excitation energy and several excited multipolarites.  (Note
    that the interval between the first three frames is about 5 times the
    interval between the lower frames.)}
\end{figure}

The dynamics described by Eqs.~\eqref{eq:tdwfs_full} is in spirit
similar to the constrained density TDHF approach developed by
\textcite{Umar:2006, Oberacker:2015}.  The nuclear systems will
evolve in time, while only the collective degrees of freedom are
active, in exactly the same manner as the theory of Adiabatic TDHF
(ATDHF) was envisioned~\cite{Baranger:1978, Brink:1976, Ring:2004}.
The present approach is formulated directly in terms of (all) relevant
collective degrees of freedom and there is no difficulty in separating
the degrees of freedom into intrinsic and collective. The definition
of the collective degrees of freedom and their proper separation from
the intrinsic ones is a problem practically unsurmountable in the
usual theory of LACM~\cite{Dang:2000}. The dynamics described by
Eqs.~\eqref{eq:tdwfs_full}, or their simplified form
Eqs.~\eqref{eq:psi}, is by default isentropic.  There is no need and
no difficulty in constructing either the inertia tensor or the
potential energy surface, as the system naturally evolves from one
point to another in the collective phase space.

\subsection{Induced Symmetric and Asymmetric Fission}

In Fig.~\ref{fig:fission} we show the evolution of a $^{238}$U nucleus starting
from the ground state with two different initial isoscalar velocity kicks and no
shell-corrections, leading to
symmetric and asymmetric fission respectively. The fragments emerge with a
significant excitation energy and deformation, clearly not in the
corresponding ground states. The
excitation energy of the nucleus has two components, the collective flow given
by Eq.~\eqref{eq:Kinetic} and the deformation and Coulomb energies given by
Eq.~\eqref{eq:NEDF} respectively. The total energy is naturally conserved.  The
results presented in these figures have been obtained by solving the full
three-dimensional (3D) time-dependent equations~\eqref{eq:nlseq}, on a $64^3$
spatial lattice.  This takes about 30 minutes on a MacBook Pro using a
\textsc{matlab} program about 250 line long (including the generation of the
ground state and of the movie and related graphics).

\subsection{Giant Isovector Resonances}

Equations~\eqref{eq:nlseq} have also been used to extract the behavior
of the \glspl{GDR} as a function of the atomic mass $A$.  As a function of the atomic
mass, the \glspl{NEDF} without entrainment yield $\hbar\omega_{\text{GDR}} \approx
65\ldots 70/A^{1/3}$~\si{MeV}, which is too low (Piotr Magierski, private
communication). Even though the entrainment term plays a small role when
computing the ground state energies, it plays an important role in determining
the excitation energies of isovector modes.

Let us consider for simplicity a
nucleus with $N=Z$, in which we will neglect the Coulomb effects and the
neutron-proton mass difference $m \approx m_n \approx m_p$.  In this case in
the ground state $\rho_n(\vect{r})\equiv
\rho_p(\vect{r}) = \rho_0(\vect{r})/2 = \abs{\psi_0(\vect{r})}^2$
and the density are determined form
the nonlinear equation
\begin{gather}
 - \frac{\tilde{\hbar}^2}{2m_\tau}\laplacian\psi_0 +
    \mathcal{E}_0' (\rho_0)\psi_0
  = \mu_0\psi_0.
\end{gather}
Considering small amplitude isovector modes $\rho_{n,p}(\vect{r},t) =
\abs{\psi_0(\vect{r}) \pm \psi_1 ((\vect{r},t) }^2$, where
$\psi_0(\vect{r})=\psi_0^*(\vect{r})$, one can derive the equation:
\begin{gather}
  i\tilde{\hbar} \dot{\psi}_1 = -(\eta+\alpha) \frac{ {\hbar}^2 }{ 2m}\laplacian
  \psi_1
  +2\mathcal{E}_0' (\rho_0) \psi_1 \nonumber \\
  + 8\frac{ \mathcal{E}_1(\rho_0) }{ \rho_0} (\psi_1+\psi_1^*) - \alpha
  \frac{{\hbar}^2}{m} \psi_0 \grad \cdot \left ( \frac{\grad
      {(\psi_0\psi_1)}}{\psi_0^2}
  \right).
\end{gather}
For $\alpha>0$ the effective mass is lowered (first term) and the stiffness
increased (last term), which results in an increased value of the
$\hbar\omega_{\text{GDR}}$. Thus, the entrainment term through the parameter $\alpha$
provides a direct handle on isovector dynamics without significantly impacting
the static properties of the \glspl{NEDF}.  One last remark: Eqs.~\eqref{eq:nlseq} will
only provide a good description of the collective modes and their corresponding
transition strengths.

\glsreset{NEDF}
\section{Conclusions}
The \gls{NEDF} developed in this work contains at most 7 significant
parameters, each clearly related to specific properties of nuclei.  In
particular, to globally fit the measured ground state energies of 2375
nuclei, functionals NEDF-En and NEDF-Enr can achieve an \gls{rms}
residual of $\chi_E\approx $~\SIrange{2.6}{2.7}{MeV} \emph{with only 4
  significant fit parameters} (and an insignificant pairing
parameter $\delta$).

Therefore NEDF-En and NEDF-Enr reproduce the behavior of the best
nuclear mass formula, with fewer free parameters, and yet have many
advantages; unlike nuclear mass formulas, these \glspl{NEDF} also model
the neutron and proton densities in the nuclei, allowing one to
extract the charge radii.  Fitting the masses alone gives an
\gls{rms} residual of $\chi_r=\SI{0.13}{fm}$ for the corresponding 883
measured charge radii, while including them in the fit improves this
to $\chi_r=\SI{0.05}{fm}$ without significantly affecting the quality
of the mass fits.

As in the mass formulas Eqs.~\eqref{eq:Bethe} and~\eqref{eq:masses},
one needs two parameters $b_0$ and $c_0$, which control the isoscalar
nuclear properties and thus are needed to reproduce the symmetric
nuclear binding energy and density.  The isoscalar nuclear
compressibility acquires a very reasonable value too, although the
saturation density is a little low, but not well constrained by the
mass fits alone.

Two other parameters $a_1$ and $b_1$ control the symmetry properties
of nuclear matter, and are correlated with a gradient term with
parameter $\eta$ that controls the diffuseness of the nuclear surface.
These are similar to the parameters $a_I$ and $a'_I$ in nuclear mass
formula Eq.~\eqref{eq:masses}, but we find that the linear combination
$a_1 - b_1\rho_0^{1/3}$ is poorly constrained by the masses.  This
gives one an essentially independent control of the ``local'' symmetry
energy slope $L_2$ (not the full $L$, which is fully determined by the neutron
equation of state), along with neutron skin thicknesses.

The addition of quartic isovector coefficients $\beta^4$ permits the NEDF
to match the neutron matter equation of state without significantly
affecting the global mass fit.  We thus find that the masses and the
neutron matter equation of state are essentially uncoupled.

These \gls{NEDF} are suitable to study complex collective motion such
nuclear fission, low energy collision of nuclei, their fusion,
interactions with various external time-dependent probes, structure
and dynamics of the neutron star crust, nucleosynthesis, etc. in a
computationally efficient manner in order to better gain intuition and
understanding about nuclear dynamics. Finally, an entrainment term
with parameter $\alpha$ controls the energies of the isovector nuclear
excitation modes and together with an additional parameter $\xi$
one can control the value of the Thomas-Reiche-Kuhn sum rule.

This \gls{NEDF} contains several new elements with respect to commonly used
Skyrme-like density functional theories:
\begin{itemize}

\item Terms proportional to $\rho^{5/3}$, similar to those found in
  the study of the unitary Fermi gas.

\item The gradient terms in Eq.~\eqref{eq:NEDF} proportional to $\eta
  \abs{\grad \rho_\tau^{1/2}}^2 = \eta \abs{\grad \rho_\tau}^2/4\rho_\tau$ or
  to $\alpha \grad{\rho_n^{1/2}}\cdot \grad{\rho_p^{1/2}}$ .

\item There is a need to consider quartic terms in isospin density
  $(\rho_n-\rho_p)^4$ in the \gls{NEDF} if one aims to describe
  correctly both nuclei and neutron matter within the same unified
  framework, and in particular the neutron star crust.
  The binding energies, charge radii, and neutron skin thickness
  appear to be insensitive to the
  properties of the neutron equation of state, which can essentially
  be fit independently.  The position of the neutron dripline appears
  to be controlled by the full
  symmetry energy  $S$ and its density dependence $L$, unlike
  neutron skin thickness, which is controlled by the ``local" density
  dependence of the symmetry energy $L_2$.
  The properties of nuclear matter in stellar
  environments (when $N\gg Z$) will therefore be controlled by
  $S$ and $L$, influencing
  for example the reaction flow in the r-process, the structure of the
  neutron star crust, and the vortex pinning mechanism in neutron star
  crust.

\item Entrainment terms $\alpha \vect{v}_n\cdot \vect{v}_p$ do not
  appear in any standard theory of large amplitude collective motion
  in nuclear physics~\cite{Baranger:1978, Brink:1976, Ring:2004},
  despite being allowed by symmetry.  They have direct analogues in
  superfluid mixtures -- the Andreev-Bashkin effect -- and are as
  natural to consider in the presence of mixed proton and neutron
  superfluids in neutron stars as they are in mixtures of \ce{^3He}
  and \ce{^4He} superfluids~\cite{Andreev:1975, Volovik:1975a,
    Vardanyan:1981}.  These terms have little influence of the ground
  state binding energies, but a strong effect on excited isovector
  modes and such terms have been discussed in the physics of neutron
  stars and determined to be relevant for the description of
  glitches~\cite{Alpar:1984x, Babaev:2004, Gusakov:2005, Chamel:2006,
    Chamel:2013a}.
\end{itemize}
To extend the accuracy of this \gls{NEDF} to reproduce finer details
of nuclear binding energies (screening of the Coulomb interaction and
a correct treatment of the exchange term, isospin and charge symmetry
breaking effects, Wigner energy, effects due to the restoration of
various other broken symmetries, etc.) first requires an accurate
accounting of shell effects; the procedure is outlined.  Next, the
parameters $\alpha$ and $\xi$ can be determined by fixing the correct value of
the Thomas-Reiche-Kuhn sum rule.
Additional parameters should be introduced into the \gls{NEDF} to model
the appropriate physics such as spin-orbit interactions and pairing.
The NEDF model is physically intuitive, providing a framework for
systematically adding physically motivated parameters.

The ability to study dynamical phenomena is a great advantage of the
present approach. It is not difficult to develop a generalization of
the NEDF to finite temperatures and to include dissipation in the
dynamics. In particular it would be extremely interesting to establish
the general trends in which mass and charge asymmetry influences the
excitation energies and average angular momentum of fragments in
induced fission as a function of the initial energy injected into the
nucleus and impact parameters in case of reactions. It would be
interesting as well to study the sticking
probability of two colliding heavy-ions as a function of the mass and
charge ratio, impact parameter and relative velocity.

A great advantage of the present approach is to provide a clear
strategy for improving the quality of NEDFs by separating the various
energy scales. Here we have identified the bulk properties, and shown
that they can be properly accounted for with a minimal number of
parameters.  Now we have a clear path forward to refine the structure
of the NEDF to properly account for smaller corrections.  In this
respect the approach outlined here is similar in spirit to an
effective field theory. The next step is to introduce the spin-orbit
coupling, the pairing coupling constants (about which there is already
an abundance of information), and to pinpoint the values of the
couplings $\alpha$ and $\xi$, all of which have a very small effect on the
nuclear bulk properties. The major impact of these new constants will
be on the shell-correction energies, and extensive past experience
indicates that the rms energy should be reduced drastically from about
\SI{2.6}{MeV} to close to \SI{0.5}{MeV} or so~\cite{ Strutinsky:1966,
  Strutinsky:1967, Strutinsky:1968, Strutinsky:1976, Brack:1972,
  Brack:1985, Myers:1966, Myers:1969, Myers:1974, Myers:1990,
  Myers:1991, Myers:1996, Ring:2004, Bohr:1998, Moller:1995,
  Moller:2012} or even below, at which point quantum
chaos will likely need to be properly accounted for~\cite{Bohigas:2002, *Bohigas:2002E, Aberg:2002,
  Olofsson:2006, Olofsson:2008, Molinari:2004, Molinari:2006,
  Hirsch:2004, Hirsch:2005, Barea:2005}.

\begin{acknowledgments}
  We are grateful to George F. Bertsch for numerous discussions and
  suggestions, to Sanjay Reddy and David B. Kaplan, who urged us to
  explore the role of $\grad{\rho_n}\cdot \grad{\rho_p}$ types of
  terms, to Rebecca Surman for discussions about aspects of neutron
  star physics, and to Jeremy W. Holt for valuable comments.  This
  work was supported in part by US DOE Grant No.\ DE-FG02-97ER-41014.
  Some calculations reported here have been performed at the
  University of Washington Hyak cluster funded by the NSF MRI Grant
  No.\@ PHY-0922770.
\end{acknowledgments}

\section{Supplemental material}

Tables with the results of the numerical fits of various \gls{NEDF}s
presented in Table~\ref{table:NEDF_A}, along with the Python code used
to perform these fits and a \textsc{matlab} code used to determine the ground
state and the 3D time-dependent fission process illustrated in
Fig~\ref{fig:fission} can be downloaded from
\url{http://faculty.washington.edu/bulgac/NEDF}.


\providecommand{\selectlanguage}[1]{}
\renewcommand{\selectlanguage}[1]{}
\bibliography{master,local}

\end{document}